\documentclass[sn-mathphys,Numbered,iicol]{sn-jnl}

\usepackage{graphicx}%
\usepackage{multirow}%
\usepackage{amsmath,amssymb,amsfonts}%
\usepackage{amsthm}%
\usepackage{mathrsfs}%
\usepackage{xcolor}%
\usepackage{textcomp}%
\usepackage{manyfoot}%
\usepackage{booktabs}%
\usepackage{algorithm}%
\usepackage{algorithmicx}%
\usepackage{algpseudocode}%
\usepackage{listings}%
\usepackage{threeparttable} 
\usepackage{placeins} 

\usepackage{tabularx}  
\usepackage{enumitem}
\newlist{mylist}{itemize}{1}
\setlist[mylist]{label=\textbullet,
                 nosep, wide, leftmargin=*,
                 before=\vspace{-0.57\baselineskip},
                 after =\vspace{-0.85\baselineskip}}
\usepackage{caption}
\usepackage{subcaption} 

\usepackage[acronym]{glossaries}
\usepackage{glossary-mcols}
\usepackage{array}
\usepackage{supertabular}

\newlength\descwidth
\newcounter{rowcount}

\newglossarystyle{table}%
{%
   {%
    \setcounter{rowcount}{0}%
    \begin{supertabular}[H]{l>{\raggedright}p{\descwidth}}%
   }
   {\end{supertabular}}%
  \renewcommand*{\glsgroupheading}[1]{}
  \renewcommand{\glossentry}[2]{%
    \stepcounter{rowcount}%
    \glsentryitem{##1}\glstarget{##1}{\glossentryname{##1}} &
    \glossentrydesc{##1}
    \ifnum\value{rowcount}=25 
      \def\next{%
        \end{supertabular}\hfill
        \setcounter{rowcount}{0}%
        \begin{supertabular}[H]{l>{\raggedright}p{\descwidth}}%
        \bfseries \entryname & \bfseries
        \descriptionname \tabularnewline
      }%
    \else
      \def\next{\tabularnewline}%
    \fi
    \next
  }%
}

\makenoidxglossaries

\usepackage{orcidlink}

\newacronym{easa}{EASA}{European Union Aviation Safety Agency}
\newacronym{arp}{ARP}{Aerospace Recommended Practice}
\newacronym{evtol}{eVTOL}{electric Vertical Takeoff and Landing}
\newacronym{etp}{ETP}{Equal Time Point}
\newacronym{fcc}{FCC}{Flight Control Computer}
\newacronym{fha}{FHA}{Functional Hazard Analysis}
\newacronym{fta}{FTA}{Fault Tree Analysis}
\newacronym{fdal}{FDAL}{Function Development Assurance Level}
\newacronym{mtom}{MTOM}{Maximum Takeoff Mass}
\newacronym{uas}{UAS}{Unmanned Aerial System}
\newacronym{uam}{UAM}{Urban Air Mobility}
\newacronym{vtol}{VTOL}{Vertical Takeoff and Landing}
\newacronym{easascvtol}{EASA SC-VTOL}{Special Condition for Vertical Takeoff and Landing Vehicle defined by the EASA}
\newacronym{easasce19}{EASA SC E-19}{Special Condition for Electric / Hybrid Propulsion System}
\newacronym{nprd}{NPRD}{Nonelectronic Parts Reliability Data}
\newacronym{fr}{FR}{Failure Rate}
\newacronym{pasa}{PASA}{Preliminary Aircraft Safety Assessment}
\newacronym{pssa}{PSSA}{Preliminary System Safety Assessment}
\newacronym{pmsm}{PMSM}{Permanent Magnet Synchronous Motor}

\theoremstyle{thmstyleone}%
\theoremstyle{thmstyletwo}%

\theoremstyle{thmstylethree}%

\begin{document}

\title[Article Title]{Battery-Electric Powertrain System Design for the HorizonUAM Multirotor Air Taxi Concept}


\author*[1]{\fnm{Florian} \sur{Jäger} \orcidlink{/0009-0002-9452-2792}}\email{florian.jaeger@dlr.de}

\author[1]{\fnm{Oliver} \sur{Bertram} \orcidlink{/0000-0002-7732-9280}}


\author[1]{\fnm{Sascha M.} \sur{Lübbe}}

\author[1]{\fnm{Alexander H.} \sur{Bismark}}

\author[1]{\fnm{Jan} \sur{Rosenberg}}

\author[1]{\fnm{Lukas} \sur{Bartscht}}

\affil[1]{\orgdiv{DLR Institute of Flight Systems}, \orgname{German Aerospace Center}, \orgaddress{\street{Lilienthalplatz 7}, \city{Braunschweig}, \postcode{38108},  \country{Germany}}}


\abstract{The work presented herein has been conducted within the DLR internal research project HorizonUAM, which encompasses research within numerous areas related to urban air mobility. One of the project goals was to develop a safe and certifiable onboard system concept. This paper aims to present the conceptual propulsion system architecture design for an all-electric battery-powered multirotor \acrfull{evtol} vehicle.
Therefore, a conceptual design method was developed that provides a structured approach for designing the safe multirotor propulsion architecture. Based on the concept of operation the powertrain system was initially predefined, iteratively refined based on the safety assessment and validated through component sizing and simulations. The analysis was conducted within three system groups that were developed in parallel: the drivetrain, the energy supply and the thermal management system.
The design process indicated that a pure quadcopter propulsion system can merely be designed reasonably for meeting the \acrfull{easa} reliability specifications. By adding two push propellers and implementing numerous safety as well as passivation measures the reliability specifications defined by \acrshort{easa} could finally be fulfilled. The subsequent system simulations also verified that the system architecture is capable of meeting the requirements of the vehicle concept of operations.  
However, further work is required to extend the safety analysis to additional system components as the thermal management system or the battery management system and to reduce propulsion system weight.
}




\maketitle

\FloatBarrier
\printnoidxglossary
\printnoidxglossary[type=\acronymtype,style=table, title={Nomenclature}]

\section{Introduction}\label{sec:Introduction}
As the majority - over $54\; \%$ (4.5 billion people) - of the world population is already living in urban areas and this trend is projected to continue up to $68\; \%$ (6.68 billion people)  by 2050, new, efficient and zero-emission transportation methods will play a major role in organizing the future mobility needs \cite{OurWorldInData.Urbanization}.  Already today, people within the biggest cities experience traffic delays of more than 100 h per person within a year \cite{INRIX.2022}. Since the ongoing urbanization will most probably lead to further congestion while the road infrastructure cannot be increased unlimited, new alternatives to existing travel methods need to be established. The vision of \acrfull{vtol} vehicle operation is to relocate some of the travel from the road up into the air and, thereby, contribute in reducing traffic jams, improving the mobility and reducing the personal travel time \cite{Goyal.2018,Rothfeld.2021} for a portion of today's travel demand.  
Another major challenge in the ongoing trend of urbanization is that today $45\;\%$ of all CO2 emissions from global transportation is produced by the road traffic \cite{OurWorldInData.Energy}. Using electrified propulsion systems for \acrshort{vtol} vehicle may contribute in providing low-emission transportation means especially when using renewable energy sources \cite{Goyal.2018,Andre.2019}. 
Lastly and more importantly, concepts for electrified \acrshort{vtol} vehicle can pave the way towards electric powered aircraft as they may be one of the first suitable use cases so far. $30.6\; \%$ of all flights within Europe cover a flight distance of less than 500 km \cite{Eurocontrol.2021}. Those flights may offer the chance to be conducted by electric aircraft \cite{Baumeister.2020, Staack.2021}.

Due to these challenges in transportation and the potentials for \acrshort{vtol} vehicles several hundred vehicle concepts have been unveiled and are worked on by aircraft manufacturers, start-ups, automotive manufacturers as well as research facilities \cite{ElectricVTOLNews.2023}. The first certified operation with passengers on board is expected to take place already in the mid-2020s \cite{Cohen.2021}. 
One of the main challenges preventing \acrshort{vtol} vehicle with passengers on board being operated today is the unproven safety and reliability of those concepts as well as missing certification standards. Within the last four years the \acrfull{easa} has established rules, named the Special Condition for \acrshort{vtol} vehicle (\acrshort{easascvtol}), as well as corresponding means of compliance \cite{EASA.2019.AMC}. The safety objectives stated therein for the category enhanced aircraft\footnote{Vehicles that are transporting passengers over congested areas fall into the category enhanced of the \acrshort{easascvtol} \cite{EASA.2019.AMC}.} set high standards similar to commercial aviation. For example, the vehicle must be able for a continued safe flight and landing even if any single system failure or a combination of failures that are not classified as catastrophic occur. Additionally, a catastrophic failure condition must have a failure rate of less than \(10^{-9}\) failures per flight hour and must not result from a single failure. As a reliable propulsion system is crucial for a safe \acrshort{vtol} operation, the design of the vehicle's propulsion system takes over an important role. So far, there has only been little research focus on analyzing the propulsion system reliability and its effects on the safe vehicle operation. However, the research that has already been conducted, indicates that the EASA safety objectives are especially difficult to meet for wingless multirotor concept vehicles \cite{Darmstadt.2019,Darmstadt.2021,Liscouet.2022}. 
This work, therefore, addresses these challenges and aims to present the conceptual design process of developing a safe propulsion system for multirotor \acrfull{evtol} vehicles as well as its implications when being applied to an exemplary use case. The goal is to provide further insights into the conceptual design to facilitate the development of the propulsion system architecture and its certification for similar \acrshort{evtol} vehicle. The research presented herein has been conducted by the Safety-Critical Systems and Systems Engineering department of the DLR Institute of Flight Systems within the DLR internal project HorizonUAM.

\subsection{Research questions  and methodological approach}\label{sec:ResearchQuestionMethod}
In contribution to the aim of this work, the following research questions will be covered: 
\begin{enumerate}
\item How should the conceptual design process of the propulsion system be carried out for an all-electric multirotor \acrshort{vtol} vehicle that is transporting passengers over congested areas so that the safety goals of \acrshort{easascvtol} can be met?
\item	What is the impact of the \acrshort{easascvtol} reliability requirements on the conceptual design of a multirotor propulsion system? 
\item Which implications does an all-electric battery-powered  \acrshort{evtol}  have  on  the propulsion system architecture besides the safety requirements?
\item Which requirements must be met by a thermal management system of the developed all-electric multirotor propulsion system? 
\end{enumerate}
To address these research questions, chapter \ref{sec:Method} describes the methodological approach for the conceptual design of a safe propulsion system for a quadcopter \acrshort{evtol} vehicle. In chapter \ref{sec:CaseStudy}, the methodological approach is applied using an exemplary \acrshort{evtol} use case of the HorizonUAM project. Up to section \ref{sec:3.ConceptDefinition}, an initial propulsion system concept is developed, which is further detailed and refined within section \ref{sec:3.Safety} based on the safety design process. For the derived propulsion system architecture, the power and drive system, the thermal management system (TMS) and the electrical system, are then sized and simulated and final architecture adjustments deducted within sections \ref{sec:3.SizingPowerSystem},  \ref{sec:3.SizingElectricalSystem} and \ref{sec:3.SizingTMSSystem}. The chapter \ref{sec:3.FinalArchitecture} presents the final propulsion architecture. Within chapter \ref{sec:Conclusion} the main findings are summarized and the initial research questions answered. The paper is completed by deriving current limitations of the applied methodology and giving an outlook for further research within chapter \ref{sec:FuturePerspective}. 

\subsection{State of the Art}\label{sec:StateOfArt}
At first, a literature research was conducted to identify the current state of the art regarding conceptual design methodologies for developing the propulsion system for multirotor vehicles, that also take safety requirements into account. Only little literature could be found that addresses this research area so far.

\vspace{5mm} 
\textbf{Conceptual design methods for the multirotor propulsion systems}\\
In 2021, \citet{Bertram.UAMDesign} developed a sizing loop which supports the initial multirotor vehicle sizing process based on flight mission requirements and the propulsion technology to be used. However, this method does not provide any detailed information about the propulsion system architecture design and its reliability or failure probabilities. 
In the work of \citet{Liscouet.2022} from 2022, a method for the conceptual design of multirotors is presented which includes a controllability analysis, a sizing optimization as well as a safety assessment. However, the controllability analysis does not take flight phase transitions and handling quality aspects into account and so far, the applicability of the safety assessment was only shown based on the \acrfull{uas} regulations. The applicability of this approach for manned \acrshort{evtol} vehicles therefore is still due.  

\vspace{5mm} 
\textbf{Currently achieved failure rates of multirotor \acrshort{evtol} vehicle architectures}\\
In 2019, \citet{Darmstadt.2019} conducted several safety analyses for the propulsion systems of in total four \acrshort{vtol} configurations, including a tilt-wing, quadcopter, lateral-twin and lift \& cruise configuration. For all developed propulsion system architectures a failure rate in the magnitude of \(10^{-4}\) per flight hour was identified, with the quadcopter configuration having the highest failure rate of \(7.97 \cdot10^{-4}\) per hour. The major challenges for multirotors in meeting the \acrshort{easascvtol} are that “a single failure must not have a catastrophic effect upon the aircraft" (VTOL.2550) and that a catastrophic event must not happen more often than once every \(10^{9}\) flight hours \cite{EASA.2019.AMC}.
The work of \citet{Liscouet.2022} also came to the conclusion that their unmanned quadcopter failure rate lies in the magnitude of \(1.44 \cdot 10^{-4}\) per hour and can effectively be reduced by adding more rotors. By using at least eight rotors – which implies a coaxial quadcopter or octocopter configuration - the \acrshort{easascvtol} could be fulfilled according to their studies. 

In 2021, \citet{Darmstadt.2021} renewed the propulsion architectures from 2019 focusing on the challenging multirotor configurations to improve their reliability and additionally expanded their safety assessment. The failure probability for the quadcopter configuration experiencing catastrophic events could be improved to \(1.78 \cdot10^{-9}\) per hour when using cross-shafts that connect all four main rotor drives. Without using a cross-shaft solution the highest failure probability increases up to \(1.75 \cdot10^{-5}\) per hour. Only by adding numerous redundancies the failure probability could be reduced to \(1.06 \cdot10^{-9}\) per hour which may be a suitable solution. Therein, the thermal management system was identified as a critical supplementary system, which needed to be dual redundant. However, the consequences of adding these redundancies on vehicle mass, complexity and feasibility of the design haven’t been further analysed. 
The difficulty of meeting the \acrshort{easascvtol} reliability goals shows that a new approach is needed that integrates the safety and reliability assessment into the conceptual design. Additionally, the implications of a safe multirotor propulsion architecture on the vehicle design, mass, the feasibility and complexity need to be readily analysed within the approach to show whether or not the system architecture should be pursued.  

The rules and regulations that must be considered within the design process are the already mentioned \acrshort{easascvtol}, the SC E-19 that define the special condition for electric or hybrid propulsion systems intended for \acrshort{vtol} aircraft, the \acrfull{arp} 4754A and \acrshort{arp}4761 that define certification considerations and safety assessment guidelines \cite{EASA.2019.AMC, ARP4761, ARP.4754A, EASA.SCE19}.\footnote{For completeness it is noted, that the CS-P define the certification specifications for the propellers and should also be taken into consideration during the propulsion architecture design \cite{EASA.2003}. However, they are out of scope of this work.}  

Therefore, this work will at first present a methodological approach which was applied for the conceptual design of an all-electric propulsion system for a quadcopter that aims at fulfilling \acrshort{easascvtol}, takes into consideration \acrshort{arp}4761 as well as the \acrfull{easasce19} \cite{EASA.SCE19} and further analyses its consequences on critical vehicle and flight mission parameters.

\section{Method}\label{sec:Method}
\begin{figure*}[htpb]
    \centering
    \includegraphics{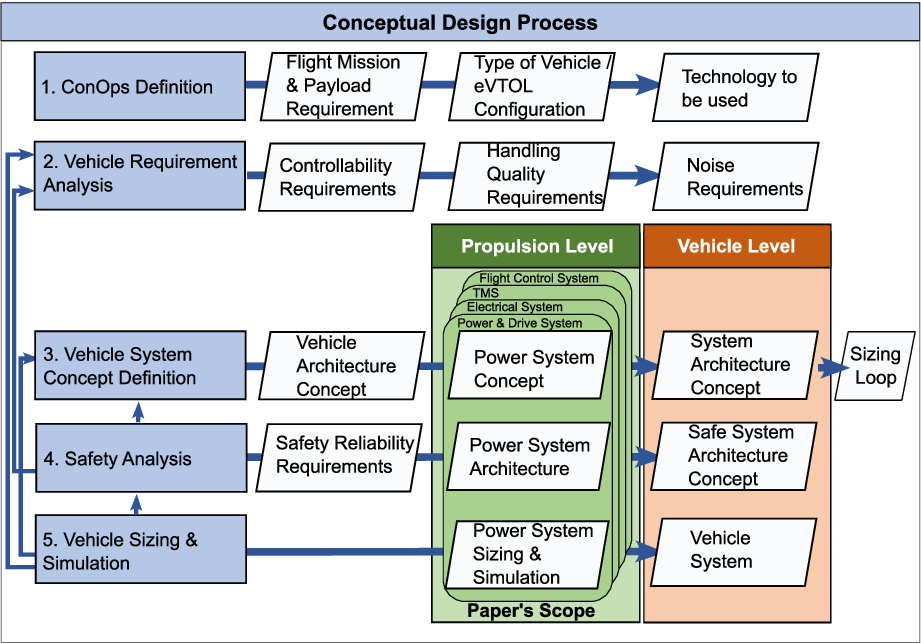}
    \caption{Applied methodological approach for the conceptual propulsion system design}
    \label{fig:method}
\end{figure*}

The applied methodological approach for designing a propulsion system consists of following steps as indicated in Fig. \ref{fig:method}:   
\begin{enumerate}
\item ConOps Definition
\item Vehicle Requirement Analysis
\item Vehicle System Concept Definition
\item Safety Analysis
\item Vehicle Sizing \& Simulation
\end{enumerate}

In the first step, the concept of operations (ConOps), the flight mission as well as the payload requirements are defined. Based on this information the best suitable type of vehicle is preselected. Thereafter, the propulsion technology for the selected vehicle is defined. 

In the second step, further design requirements for the propulsion system and its components are derived. Several topics should be included herein, among them controllability, handling quality and noise aspects, as they may heavily influence the propulsion system design.
  
The ConOps definition and the requirement analysis are the basis for step three, the propulsion system design loop. Within this step, an initial system concept consisting of the propulsion system as well as the other vehicle systems is defined. Thereby, the propulsion system is specified in terms of the power system, electrical system and the thermal management system. Based on the initial vehicle system architecture concept a first vehicle sizing loop is performed to derive vehicle parameters as, for example, the required vehicle propulsion power, the required energy, vehicle empty weight and rotor size. 

In the fourth step, a safety and reliability analysis is conducted for the initial vehicle system architecture, in order to fulfil the \acrshort{easascvtol} safety goals. Any architecture changes are then passed back to the vehicle system concept. 

Within the last step, the safe vehicle architecture is modelled and simulated to validate the suitability of the derived vehicle architecture based on the ConOps definition and vehicle requirement analysis. 

As this paper focuses on the propulsion system architecture concept, emphasis is put on presenting steps three, four and five primarily for the propulsion system. The steps one and two are only briefly described in order to provide the context for the propulsion system design.

\subsection{ConOps Definition}\label{sec:2.ConOps}
When detailing the ConOps, the type of \acrshort{evtol} vehicle needs to be selected based on the intended use case and payload requirements. As shown by \citet*{Ratei.2021} different vehicle concepts may be suitable for different operating areas. For example, a rotary-wing concept like a multirotor, or fixed-wing concepts like the lift \& cruise, or vectored thrust configurations with tilted wing, tilted rotors or tilted ducts may be suitable.  

As soon as the flight mission is defined and the \acrshort{evtol} vehicle is chosen, the propulsion technology to be used should be evaluated. Propulsion systems like a full-electric battery-powered or hydrogen powered system or even serial or parallel hybrid electric solutions may be suitable.\footnote{A comparative overview of the characteristics of different propulsion technologies used for a multirotor and their impact on the application areas are presented within \cite{Bertram.UAMDesign,Bertram.Impact}.}

\subsection{Vehicle Requirement Analysis}\label{sec:2.RequirementAnalysis}
\begin{figure}[htpb]
    \centering
    \includegraphics[scale=0.9]{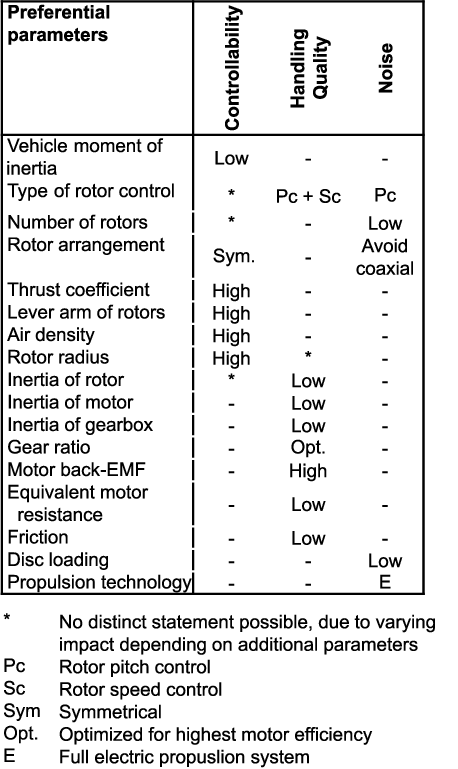}
    \caption{Overview of some selected parameters influencing controllability, handling quality and noise. The parameters may be dependent on each other and the list is not exclusive.}
    \label{tab:RequirementTable}
\end{figure}
Within step two, several further design requirements, primarily for the propulsion system design are collected. As the flight control functions are primarily taken over by the propulsion system within \acrshort{evtol} vehicles, controllability and handling quality requirements have a significant impact on the propulsion system design and should therefore already be considered at an early design stage. 
The controllability analysis aims at ensuring, that the vehicle is controllable around all axes. This requires that the Newton's law for translation and rotation need to be fulfilled. As described within \citet*{Liscouet.2022}, it is essentially not only to analyse the controllability during normal operating conditions, but also for any failure cases. The difficulty during this step is that the failure cases are not known in the beginning. Therefore, the controllability for failure cases needs to be analysed again when having identified the failure cases within step four, the safety analysis. In general, the controllability depends on whether rotor speed control or pitch control is used and is influenced by the vehicle moment of inertia, the number of rotors, the rotor arrangement, additional rotor parameters (e.g. rotor inertia) and the thrust coefficient of the corresponding rotors \cite{Pavel.2021}. 
In addition to the controllability aspects, the handling qualities should be considered. According to \citet*{Pavel.2021} the handling qualities of a multicopter are influenced not only by the aircraft response to a control input (controllability) but also to the coupled rotor-motor dynamics. The dynamic response of a coupled rotor-motor drive system, is beneficially influenced by a low  rotor inertia, low inertia of the drive components (motor and gearbox), a drive system gear ratio which is optimized for high motor efficiency, balanced motor performance\footnote{Which is expressed in terms of the ratio $\frac{K_{e}^2}{R_{a}}$ consisting of the motor back-electromotive force constant ${K_{e}^2}$ and the equivalent resistance ${R_{a}}$.}, a low motor equivalent resistance and low friction losses within the drive system which are influenced by the rotational speed and the gear ratio \cite{Pavel.2021}. All influencing parameters combined result in good handling qualities when the dynamic response around all axes is characterised by low rise-time, high bandwidth, low overshoot and high stability in terms of phase and gain margin \cite{Pavel.2021}.
According to the analysis of \citet*{Bahr.2021} and \citet*{Niemiec.2020} for rotor speed-controlled quadcopters, the weight of the electric motors that are only sized based on the maximum required power demand for performing the flight mission, is insufficient for meeting handling quality requirements. As a high motor torque capability is required for achieving a low motor time constant and therefore good handling qualities, the motor becomes twice as heavy as initially sized. The sum of all electric motors might reach $15-16\; \%$ of the total vehicle weight in case a direct drive is used.
Therefore, the factors influencing the handling qualities should already be considered within the conceptual design phase to minimize the design adjustments at a later design phase.\footnote{For more information about analysing the handling qualities of multicopter \acrshort{evtol} vehicles it is referred to \cite{Atci.2021}, \cite{Atci.2022} and \cite{Atci.2022b}.} 

When designing the propulsion architecture the noise level as well as the effect of noise annoyance should be taken into account in order ensure public acceptance \cite{Gong.2022}. The \acrshort{evtol} architecture parameters type of rotor control, maximum rotor tip speed, number of rotors, rotor arrangement, disc loading and the propulsion system architecture should be carefully chosen, as they mainly influence the emitted noise according to \citet*{Brown.2018, Smith.2020} and \citet*{Smith.2021}.

An overview of the different parameters and their optimum values is given within Fig. \ref{tab:RequirementTable}. 

\subsection{Propulsion System Concept Definition}\label{sec:2.PropulsionConceptDefinition}

Taking all the aspects of the ConOps definition and the vehicle requirements analysis into consideration a first propulsion concept is established. In this step all systems are identified that are required within the \acrshort{evtol} propulsion system. At this point it is important to differentiate between the different nomenclature that is used to describe the propulsion systems. Herein the nomenclature offered by \citet*{Herrmann.2015} is followed in which the propulsion system encompasses a group of systems that contribute to provide lift and power the \acrshort{evtol}. The propulsion concept was developed by conducting the following steps:

\begin{enumerate}
    \item At first, the propulsion system context is defined, which identifies external elements interacting with the propulsion system. The type of interaction is defined by the interfaces.
    \item Then the use cases for the propulsion system are established and the tasks of each system context element for each use case are defined. This allows to identify all tasks and functions that need to be fulfilled by the propulsion system.
    \item For each derived function an activity diagram is developed to describe the activities that are taking place within the propulsion system itself.
    \item With this information an initial system concept is derived by grouping the identified activities and allocating them to a specific system. In accordance with \citet*{Darmstadt.2021} the propulsion system architecture is generally composed of the following system groups: 
    \begin{itemize}
        \item Flight control system
        \item Power and drive system
        \item Electrical system
        \item Thermal management system (TMS)
    \end{itemize}
    The flight control system encompasses all sensors and systems that collect air data, receive and process control commands and calculate the corresponding motor control inputs for each motor controller for speed-controlled rotors or inputs for the actuation system of collective pitch-controlled rotors. The power and drive system (also called powertrain and drivetrain) is responsible for converting the electrical power, which is supplied by the electrical system, into mechanical rotational power in accordance with the inputs received from the flight control system. The drivetrain is a system group within the powertrain and does only include the systems that transmit the mechanical power of the engine into thrust at the rotors. The electrical system takes over the function to store and distribute the electrical energy. The thermal management system shall ensure to keep all system components within their operating temperature range. These system groups can provide an initial guidance during the architecture developing process.  

\item In the last step the propulsion concept is integrated into the complete vehicle system architecture concept and an initial sizing loop is conducted to size the \acrshort{evtol} vehicle as well as the propulsion system, estimate the required power and energy and calculate the estimated weight proportions using the method presented by \citet*{Bertram.UAMDesign}).
\end{enumerate}

\subsection{Safety and Reliability Analysis}\label{sec:2.SafetyAnalysis}
The basis for the safety analysis are the safety and reliability requirements of \acrshort{easascvtol} \cite{EASA.2019.AMC} and the \acrshort{easasce19} \cite{EASA.SCE19}. 
With the initial understanding about the intended propulsion system components from the previous section, the safety analysis helps to identify weak points of the propulsion system and to define the type and amount of required safety measures. Thereby, system requirements for each propulsion system component can be derived which may significantly impact the propulsion system design compared to the initial design. Consequently, a more precise prediction about the required system components and their specifications can be generated. 

\begin{figure*}[htpb]
    \centering
    \includegraphics{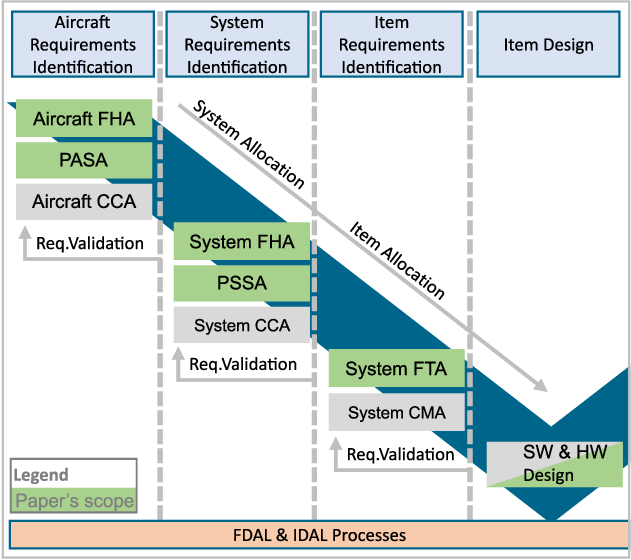}
    \caption{Extract of the ARP4754A \cite{ARP.4754A} and illustration of the safety assessment steps covered within this work (marked in green)}
    \label{fig:SafetyLoop}
\end{figure*}

Generally, the safety analysis for the propulsion system design is conducted using the methods described in SAE \acrshort{arp}4754A \cite{ARP.4754A} and \acrshort{arp}4761 \cite{ARP4761}. In Fig. \ref{fig:SafetyLoop} an extract of the safety assessment process is shown. 
The green parts mark the steps of the safety assessment that are covered within this work. In order to conduct the aircraft level \acrfull{fha} the system concept from the previous section as well as a functional breakdown analysis on the aircraft level are required, which are assigned to the system development process within the \acrshort{arp}4754A. However, as this aircraft level functional analysis has yet not been addressed, it is herein conducted under the topic of the safety analysis. Thereafter, the aircraft level \acrfull{fta}, the system level \acrfull{fha} and \acrshort{fta} are conducted while iteratively gathering the information for the \acrfull{pssa} and \acrfull{pasa} during the system design adjustments. During each iteration of the design process, the granularity of the considered systems within the safety analysis can be increased. 
Initially, a functional breakdown is being conducted to identify the main aircraft functions. In this context, especially those functions that are taken over by the propulsion system are of special interest. In the next step an \acrshort{fha} is being conducted on the aircraft level which identifies the failure cases of the previous functions and their effects on, for example, the aircraft, the passengers, the vehicle and the environment. As proposed in \citet*{Schafer} the failure cases \textit{total loss of function}, \textit{partial loss of function}, \textit{unannunciated loss of function}, \textit{incorrect operation of function}, \textit{inadvertent operation of function} and \textit{unable to stop the function} should be analysed and their failure effects on the aircraft be described. 

\begin{figure*}[h!]
    \centering
    \includegraphics{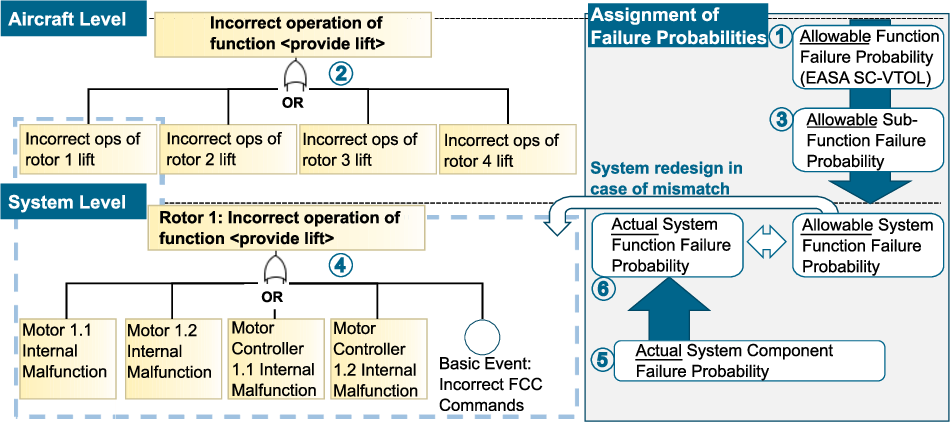}
    \caption{Schematic depiction of the interconnection between Aircraft Level FTA and System Level FTA}
    \label{fig:Aircraft_vs_SystemLevel}
\end{figure*}

The failure effect of a functional failure is the basis for the following process of developing a safe system architecture as indicated in Fig. \ref{fig:Aircraft_vs_SystemLevel}.
Based on the failure effect and the required \acrfull{fdal} as defined within the \acrshort{easascvtol} an allowable failure probability is assigned to each functional failure case (label 1). For each identified functional failure case a subsequent fault tree is created within the \acrshort{fta} on aircraft level to identify the causes (base events) that contribute to each functional failure case, the so-called top event (label 2). Based on the allowable failure probability for each failure case an allowable failure probability can be assigned to each failure cause (label 3).

Up to this point the analysis has been conducted based on the aircraft functions. In the next step the system level is analysed by identifying the systems that contribute in fulfilling a specific aircraft function. The initial system architecture from section \ref{sec:2.PropulsionConceptDefinition} is the basis for this analysis and needs to be crosschecked against all safety requirements that were derived from the aforementioned process (label 1-3). The crosscheck is done by developing a system level \acrshort{fta}, in which the top event of the system level \acrshort{fta} is the base event of the previously created aircraft level \acrshort{fta}. Within this system level \acrshort{fta} the component failure causes leading towards the top event are collected (label 4) and their failure probabilities defined by using historical data as provided by e.g. the \acrfull{nprd} Dataset \cite{Quanterion} (label 5). With this information the actual failure probability of a system function is calculated bottom-up (label 6) and gathered within the \acrshort{pasa}. As long as the allowable system reliability cannot be assured by the designed propulsion system architecture, the architecture needs to be adjusted and the process starts again. Finally, the results are collected within the \acrshort{pssa}, which indicates if the requirements of the aircraft level \acrshort{fha} can be fulfilled. 

To evaluate the sensitivity of the propulsion system architecture to critical system components, minimal cut sets are generated based on Reliability Block Diagrams (RBDs) and fed back into the system design.

As the propulsion system is a safety-critical system, whose failure may cause human injury or even loss of life, the system shall only have two states: operational or failed-safe \cite{Dubrova.2013}. A fail-unsafe condition shall be prevented by all means. Therefore, either the design principle of a safe life or fault-tolerant design must be applied for developing the propulsion system.  While a safe life design is characterized by oversizing and prematurely replacing components before failure, a fault tolerant design requires to incorporate hardware, information, time or software redundancy \cite{Dubrova.2013}. The following strategies are promoted herein to ensure a safe system design depending on the analysed aircraft functional failure case:
\begin{itemize}
\item \textbf{Total loss or partial loss of function}: \\
Implement additional system components with the same functionality (for example: passive, active or hybrid hardware redundancy).\\
\item \textbf{Unannunciated loss of function}: \\ 
Make use of software redundancy by implementing fault detection and fault indication mechanisms. \\
\item \textbf{Incorrect operation, inadvertent operation, unable to stop a function}:\\
Implement options for masking a faulty system component.\\
\end{itemize}

\subsection{Propulsion System Component Sizing and Validation}\label{sec:2.Sizing}
The last step within the applied propulsion system design methodology aims at further specifying the propulsion system components and validating the system architecture by sizing and simulating each component based on off-the-shelf components. The sizing is conducted using common sizing methods.\footnote{For detailed information about the sizing process of a multicopter propulsion system it is referred to \citet*{Bertram.2022b}.} The simulation of the propulsion system helps to validate if the derived vehicle architecture can suitably fulfil the initial ConOps definition and requirements. If necessary, additional architecture adjustments and requirements are derived based on the sizing and simulation results.

\FloatBarrier
\section{Case Study}\label{sec:CaseStudy}
Within this section, the previously described conceptual design method is applied to a case study to derive a conceptual propulsion architecture for a battery-powered multirotor \acrshort{evtol} vehicle that is operated by a pilot.

\subsection{ConOps Definition \& Vehicle Requirement Analysis}\label{sec:3.ConOps}
\begin{figure}[htpb]
    \centering
    \includegraphics{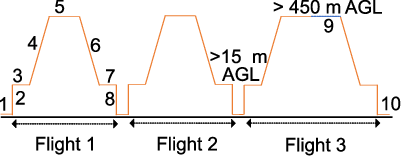}
    \caption{Flight mission profile of the analysed use case consisting of three consecutive flights with the flight segments 1: Taxi out, 2: Vertical climb to 50ft AGL, 3: Transition to cruise, 4: Cruise climb to cruise altitude, 5: Cruise until destination, 6: Cruise descent, 7: Re-transition to hover, 8: Vertical descent to ground, 9: Diversion to alternate, 10: Taxi in. \cite{Bertram.UAMDesign}}
    \label{fig:ConOps-Definition}
\end{figure}
The general requirements of the ConOps are summarized in Table \ref{tab:3.ConOps} and define the flight mission, payload, type of \acrshort{evtol} vehicle and the powertrain technology. 
\begin{table}[h]
\caption{General requirements of the ConOps definition}\label{tab:3.ConOps}%
\begin{tabular}{@{}llll@{}}
\toprule
Parameter & Value \\
\midrule
Flight Mission          &  Three Flights + 20 min Loiter       \\
Design Range            &  50 km                        \\
Payload                 &  360 kg (4 Passengers incl. Pilot)        \\
MTOM                    &  $<3175$ kg                   \\
Vehicle configuration   &  Multirotor                    \\
Powertrain technology   &  All-electric                    \\
Energy source           &  Battery                          \\
\botrule
\end{tabular}
\end{table}
For this case study the vehicle under investigation shall be operating within the intracity use case. The total flight mission, as indicated in Fig. \ref{fig:ConOps-Definition}, consists of in total 50 km, which are separated in three flights to allow for passenger embarkation and disembarkation in between plus additional 20 minutes reserve time at minimum drag speed. The \acrshort{evtol} vehicle shall be able to transport at least in total 4 passengers with 90 kg each, which results in a payload capability of 360 kg.
\begin{figure}[htpb]
    \centering
    \includegraphics{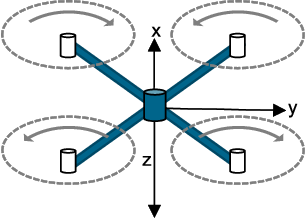}
    \caption{Quadcopter rotor cross-configuration}
    \label{fig:RotorArrangement}
\end{figure}
As up to date the quadcopter is the most critical vehicle configuration in terms of safety as described within section \ref{sec:StateOfArt} as well as energy consumption, the quadcopter vehicle configuration shall be selected for fulfilling this mission. Thereby, a cross-configuration of the rotors is selected, in which they are evenly and symmetrically distributed along the x- and y-axis to provide good controllability and handling quality as shown in Fig. \ref{fig:RotorArrangement}. The rotors shall be fixed-pitch and speed-controlled as they promise less system complexity, even though the pitch rotor control would be beneficial in terms of controllability, achieving quick vehicle response times and low noise. 
The powertrain shall be all-electric and powered by batteries, since it could be shown in \citet*{Bertram.UAMDesign} that a battery full electric powertrain is competitive compared to other powertrain technologies for up to 50 km design range, based on the current state of the art.

The controllability analysis, handling quality and noise aspects presented within section \ref{sec:2.RequirementAnalysis} have been taken into consideration in parallel to any system design adjustments. As they are not in focus of this paper, they are not further elaborated on herein.

\subsection{Propulsion System Concept Definition}\label{sec:3.ConceptDefinition}
\begin{figure*}[htpb]
    \centering
    \includegraphics{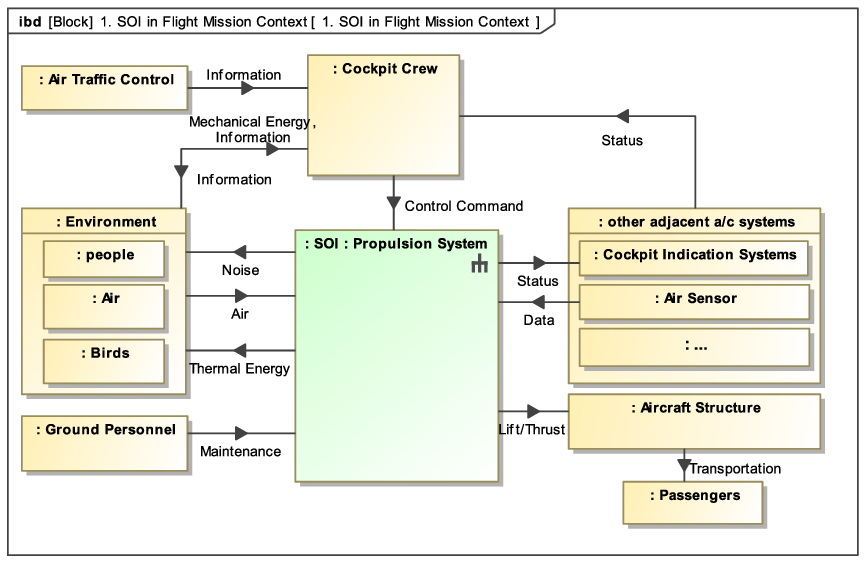}
    \caption{System context definition for the propulsion system}
    \label{fig:SystemContext}
\end{figure*}
With the information about the ConOps and the vehicle requirements an initial propulsion system concept is defined within step three of the applied conceptual design method. To develop the propulsion system concept, the CAMEO Systems Modeler was used. 
As described in section \ref{sec:2.PropulsionConceptDefinition}, initially the system context for the system of interest is defined as shown in Fig. \ref{fig:SystemContext}.

The system context indicates that the propulsion system of the multirotor receives control commands from the cockpit crew. These control commands are merged with data from the air data sensors in order to lift and control the aircraft for passenger transport. In order to provide a closed control loop for the unit controlling the vehicle, currently a cockpit crew, status information is fed back to the cockpit indication systems.  
During the transformation of the input signals into lift and thrust, air, thermal energy and noise are interchanged between the environment and the propulsion system. As the propulsion system components will suffer of degradation during daily operation, the means for maintenance actions are included within the system context.

\begin{figure*}[htpb]
    \centering
    \includegraphics[scale=0.9]{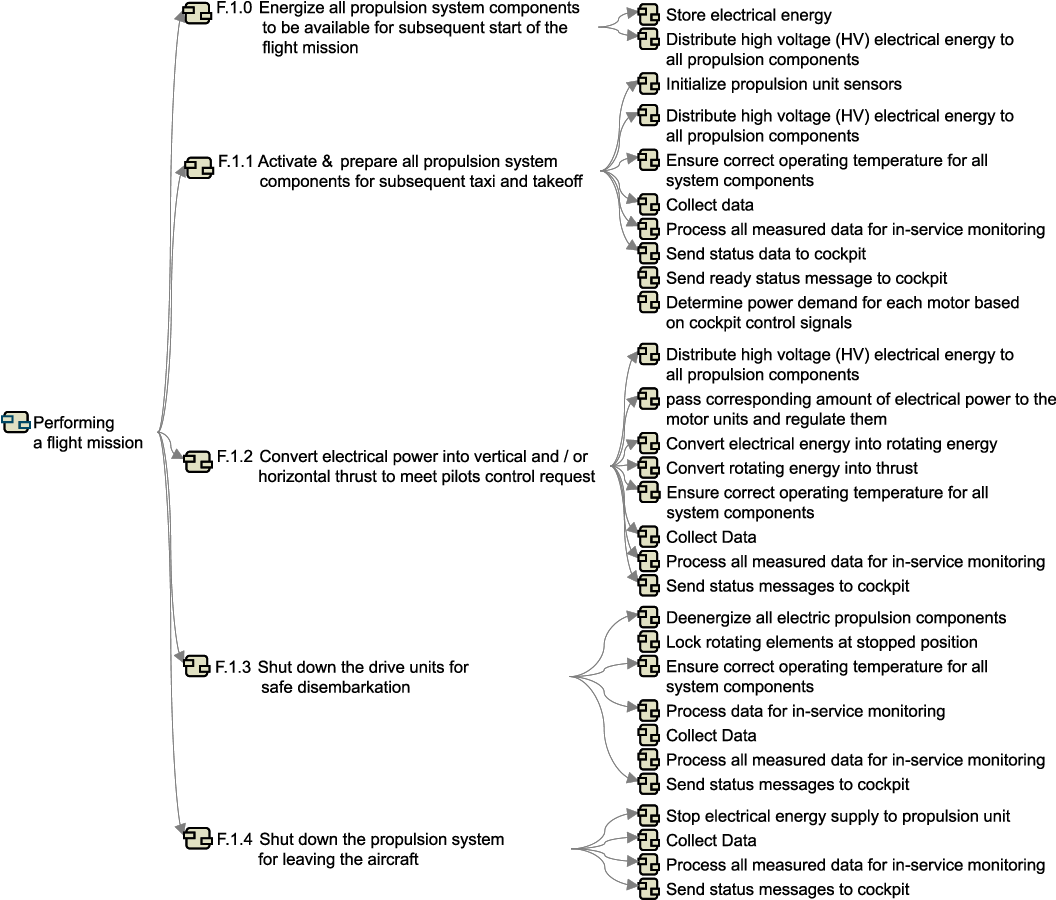}
    \caption{Propulsion system function analysis}
    \label{fig:Functions_SOI}
\end{figure*}

By analysing the use cases of the propulsion system, numerous main functions could be identified that need to be fulfilled by the propulsion system and the adjacent aircraft systems. The activity diagrams for each identified main function enabled to identify the sub-functions for the propulsion system. Fig. \ref{fig:Functions_SOI} gives an overview of all identified functions. 

\begin{figure*}[htpb]
    \centering
    \includegraphics[width=\textwidth]{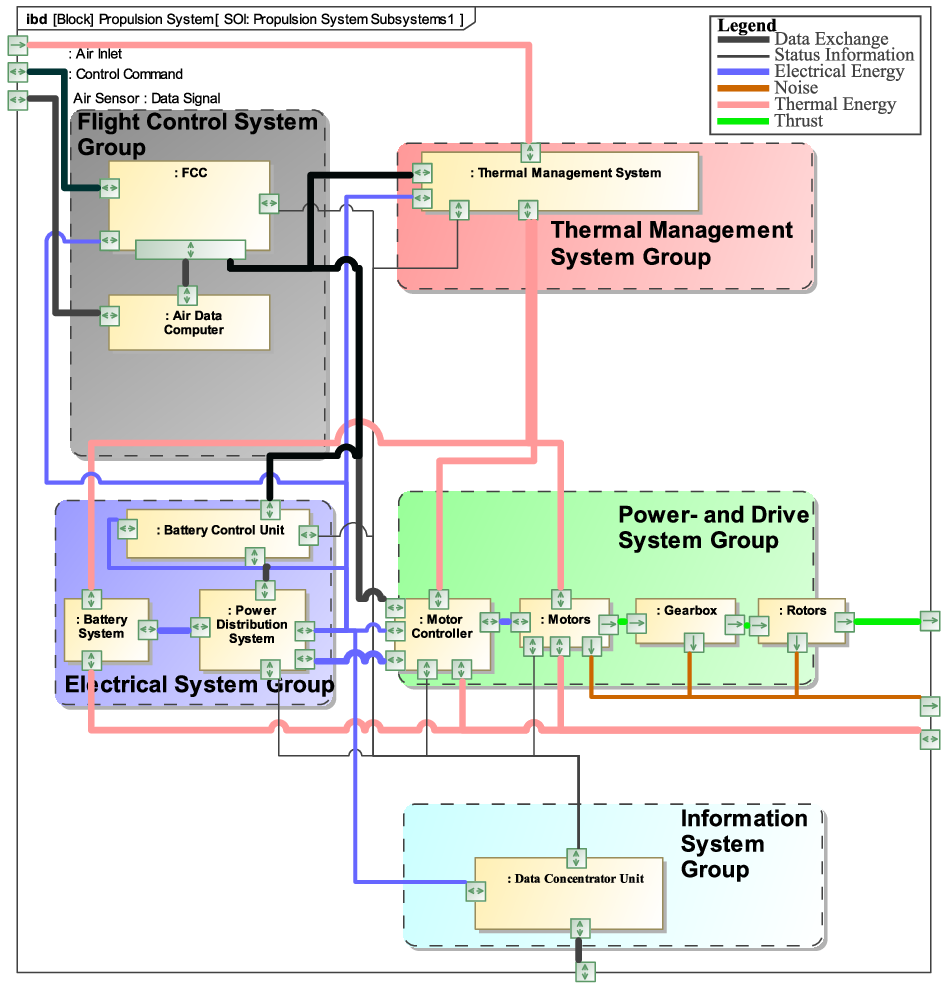}
    \caption{Logical Architecture of the Propulsion System}
    \label{fig:LogicalArchitecture}
\end{figure*}

When grouping similar functions together they can be allocated to respective system groups as shown in Fig. \ref{fig:FunctionalArchitecture}. It becomes apparent that the four generic system groups, namely the flight control, the power and drive, the electrical and the thermal management system group, as described in section \ref{sec:2.PropulsionConceptDefinition}, can be identified here as well. In addition to that, the functional analysis requires the integration of an information system group into the propulsion system as it is essential to feed back the information from all participating propulsion system groups back to the cockpit and an optional in-service monitoring unit for health monitoring and improving maintenance schedules. Besides the identification of the main system groups, this graphical representation allows to identify the item flow between the different system groups. 
Within the next step of the architecture development each functional block is assigned to a specific system component as shown in Fig. \ref{fig:LogicalArchitecture} and, thereby, an initial logical propulsion system architecture is developed. 
Within this architecture the flight control system group is composed of at least a \acrfull{fcc} and an air data computer gathering and distributing air sensor data including GPS data.  Using the control inputs from the cockpit and the air data, the \acrshort{fcc} calculates and controls the required power setting for the electric motor and, thereby, regulates also the corresponding setting of the thermal management system. The power and drive system group consists of motor controllers, motors, optionally gearboxes and the rotor. The energy for the power and drive system group is provided by electrical system group which consists of the battery system, a power distribution system and battery control units. The information system group consists of data concentrator units which gather and distribute status information. The thermal management system is not further specified, as its system requirements are unknown so far. However, based on the sizing and simulation of the power and electrical system group within section \ref{sec:3.SizingPowerSystem}, some specifications for an thermal management system are collected in section \ref{sec:3.SizingTMSSystem}. 
Merging this logical architecture with the other vehicle systems as described within section \ref{sec:2.PropulsionConceptDefinition} an initial sizing loop for the whole vehicle system is conducted. However, as this is not part of this paper, it is referred to \citep{Bertram.UAMDesign}.

Within the subsequent safety analysis, the propulsion architecture is further modified to fulfil the safety requirements of \acrshort{easascvtol} \cite{EASA.2019.AMC} of the enhanced\footnote{As soon as the vehicle is expected to transport passengers over congested areas it falls into the certification category enhanced of \acrshort{easascvtol} with the highest required safety levels.} vehicle category. The considered systems within the safety analysis are indicated in Fig. \ref{fig:LogicalArchitecture}. However, for simplification reasons, the thermal management system and the information system group are initially excluded from the safety assessment and will be integrated in the future. 

\subsection{Safety and Reliability Analysis}\label{sec:3.Safety}
The safety and reliability analysis as presented in section \ref{sec:2.SafetyAnalysis} has been conducted using the SysML Modelling Language within the CAMEO Systems Modeler together with the SysML Profile Risk Analysis and Assessment Modelling Language (RAAML) \cite{RAAML} and the \acrshort{fha} profile \cite{Schafer} which facilitate conducting a model-based safety assessment. 
The safety analysis loop has been run through several times during the design process to account for the system changes that were required to reach the reliability guidelines and to account for the results of the sizing process. 
Therefore, the following section presents the main results of each step of the safety analysis based on the final propulsion system architecture as presented in section \ref{sec:3.FinalArchitecture}. 
\vspace{5mm}

\underline {Functional Breakdown Analysis}
\begin{figure}[htpb]
    \centering
    \includegraphics{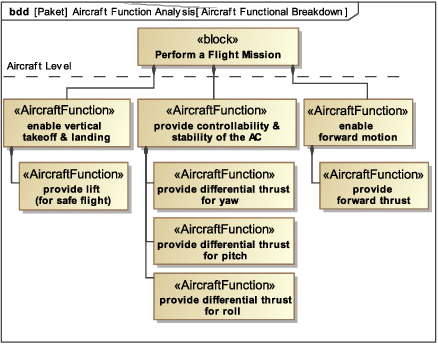}
    \caption{Functional Breakdown analysis with the aircraft functions that are connected to the propulsion system}
    \label{fig:Functional Breakdown Analysis}
\end{figure}
Within the functional breakdown analysis the main aircraft level functions were identified, that are taken over by the propulsion system as shown in Fig. \ref{fig:Functional Breakdown Analysis}:
\begin{enumerate}
    \item provide lift for safe flight, 
    \item provide differential thrust for yaw,
    \item provide differential thrust for pitch,
    \item provide differential thrust for roll and 
    \item provide forward thrust.
\end{enumerate}
\vspace{5mm}

\begin{figure}[htpb]
    \centering
    \includegraphics{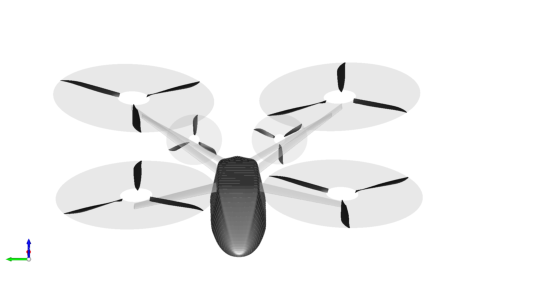}
    \caption{Visualization of the quadcopter with two push propellers \cite{Atci.2022}}
    \label{fig:VehicleVisualisation}
\end{figure}

\underline {Functional Hazard Analysis}
During the initial safety analysis loop it became quickly apparent that a quadcopter with a non-redundant propulsion system as presented in previous section\footnote{During the first iteration a quadcopter vehicle configuration is assumed in which each rotor is powered by a pure series connection of the system components \acrshort{fcc}, battery, motor controller and electric motor as described in the previous section \ref{sec:3.ConceptDefinition}.} will require so many additional redundancies that the design will most probably not be reasonable in terms of total weight and system complexity. The difficulty of the quadcopter configuration is that a partial loss of providing lift may be caused, for example, by a single rotor failure which exhibits a failure probability of $2.83 \cdot10^{-4}$ per hour\footnote{This failure rate results when using the component failure rates presented later on in Table \ref{tab:FailureRatesComponents}.}. The \textit{partial loss} of providing lift caused by a single rotor failure within a quadcopter configuration  must be expected to be a catastrophic event \cite{Darmstadt.2019} which shall not happen more often than $1.0 \cdot10^{-9}$ per hour as it is categorized as a \acrshort{fdal} A event within \acrshort{easascvtol} \cite{EASA.2019.AMC}. Therefore, a main vehicle design adjustment was conducted by adding two push propellers to the rear of the vehicle as shown in Fig. \ref{fig:VehicleVisualisation}.
This measure assumes, that the resulting yaw moment of one rotor loss can be counteracted and the failure effect of a partial loss of providing lift caused by one single main rotor loss attenuates from a catastrophic event to a hazardous event with an allowable failure probability of $1.0 \cdot10^{-7}$ per flight hour and less redundancies can be expected to be required.  

\begin{table*}[h]
\caption{FHA failure effect classification for each identified aircraft level function}\label{tab:3_FHA}
\begin{tabular}{l p{2cm} p{2cm} p{2cm} p{2cm} p{2cm}}
\toprule
Function failure    & Provide lift   & Provide diff. thrust for pitch    & Provide diff. thrust for roll    & Provide diff. thrust for yaw    & Forward thrust  \\
\midrule
Total loss          & catastrophic  & catastrophic                      & catastrophic                      & hazardous                     & major         \\
Partial loss        & hazardous     & hazardous                         & hazardous                         & n.a                           & n.a  \\
Incorrect ops.       & catastrophic  & catastrophic                      & catastrophic                      & hazardous                     & catastrophic         \\
Inadvertent ops.     & catastrophic  & catastrophic                      & catastrophic                      & hazardous                     & catastrophic          \\
Unable to stop      & catastrophic  & catastrophic                      & catastrophic                      & hazardous                     & catastrophic     \\
Unsym. partial loss & n.a.          & n.a.                               & n.a.                         & minor                         & minor        \\
Degradation         & major         & major                             & minor                             & minor                         & minor         \\
\botrule
\end{tabular}
\end{table*}

An aggregated overview of the \acrshort{fha} results based on this quadcopter configuration with two push propellers is given in Table \ref{tab:3_FHA}. Especially the failure cases \textit{total loss} or \textit{partial loss} of a function as well as the \textit{inadvertent, incorrect operation} including the \textit{unable to stop} functional failure  are critical for the system design since they exhibit catastrophic or hazardous events.  
For all functions that have at least a catastrophic and hazardous failure effect the corresponding Aircraft \acrshort{fta} is being conducted. 
\vspace{5mm}

\begin{figure}[htpb]
    \centering
    \includegraphics{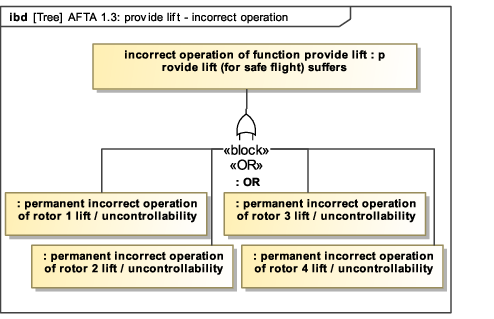}
    \caption{Aircraft FTA for the functional hazard "Incorrect operation of the function: provide lift"}
    \label{fig:AFTAIncorrectOpsLift}
\end{figure}

\underline{Aircraft Level \acrshort{fta} Results}
The aircraft \acrshort{fta} for the catastrophic event \textit{incorrect operation} of the function $<$provide lift$>$ is presented exemplarily in Fig. \ref{fig:AFTAIncorrectOpsLift}.
As the permanent incorrect operation of the function to provide lift is classified as a catastrophic event with an allowable failure probability of $1.0 \cdot10^{-9}$ per hour it can be concluded that the permanent incorrect operation of one rotor is allowed to happen with a probability lower than $2.5 \cdot10^{-10}$ per flight hour. With this information the system architecture that propels each rotor is designed in the next step using the system level \acrshort{fta}. 
By developing all \acrshort{fta}s for the catastrophic failure effects of the aircraft functions, it becomes apparent that, according to the herein used definition, the inadvertent operation or unable to stop functional failures are subsets of the incorrect operation. Therefore, three main basic events remain that need to be further analysed within the system level \acrshort{fta}: 
\begin{itemize}
    \item Total loss of one rotor lift
    \item Incorrect ops. of one rotor providing lift
    \item Incorrect or inadvertent ops. of one propeller providing thrust
\end{itemize}
\vspace{5mm}

\underline{System Level \acrshort{fta} Results}
\begin{table*}[]
\caption{Failure rate probability for each propulsion system component} \label{tab:FailureRatesComponents}
\begin{threeparttable}
\begin{tabular}{@{}llll@{}}
\toprule
& Component & Failure condition & Applied Failure Rate\footnotemark[1] \\
\midrule
BAT     & Battery                   & Failure                                   &$9.31 \cdot10^{-5}$  \cite{Quanterion}\\
MC      & Motor Controller          & Failure                                   &$4.75 \cdot10^{-5}$ \cite{Darmstadt.2019}\\
M       & Electric Motor            & Failure                                   &$9.24 \cdot10^{-5}$ \cite{Darmstadt.2019}\\
GB      & Gearbox                   & Failure                                   &$5.00\cdot10^{-6}$ \cite{Darmstadt.2019}\\
FCC     & Flight Control Computer   & Failure, Malfunction                      &$1.57 \cdot10^{-5}$ \cite{Quanterion}\\
REL     & Disconnect Power Relay    & Unintended opening, Failure to operate    &$4.60 \cdot10^{-5}$ \cite{Quanterion}\\
DISC    & Disconnect Clutch         & Unintended opening, Failure to operate    &$4.70 \cdot10^{-5}$ \cite{Quanterion}\\
\botrule
\end{tabular}
\begin{tablenotes}
  \item[1] Due to lack of data it is assumed that the applied failure rate is the same for the different cases of failure condition.
\end{tablenotes}
\end{threeparttable}
\end{table*}
On the system level, the rotor drive system architecture is revised until the \textit{total} loss and the \textit{incorrect ops} of one rotor fulfils the failure probability goals defined within the aircraft level \acrshort{fta}s of $<2.5\cdot10^{-8}$ and $<2.5\cdot10^{-10}$ respectively. Using a single drive unit without any redundancies and assuming the failure probabilities as listed in Table \ref{tab:FailureRatesComponents} the \textit{total loss} of one rotor must be expected to occur with a probability of $2.8\cdot10^{-4}$. Therefore, as a first countermeasure the rotor drive is designed as a dual active drive system composed of two drive units. As the dual active drive system would still exhibit a failure rate for a \textit{total loss} of one rotor of $5.6\cdot10^{-8}$ each motor controller unit shall be powered by at least two separate battery packs and a dual active / passive channel motor controller shall be used. The passive channel continuously monitors the main motor controller and shall be able to take over its function (hot standby redundant system).

\begin{figure*}
    \centering
    \includegraphics[scale=0.69,angle=90]{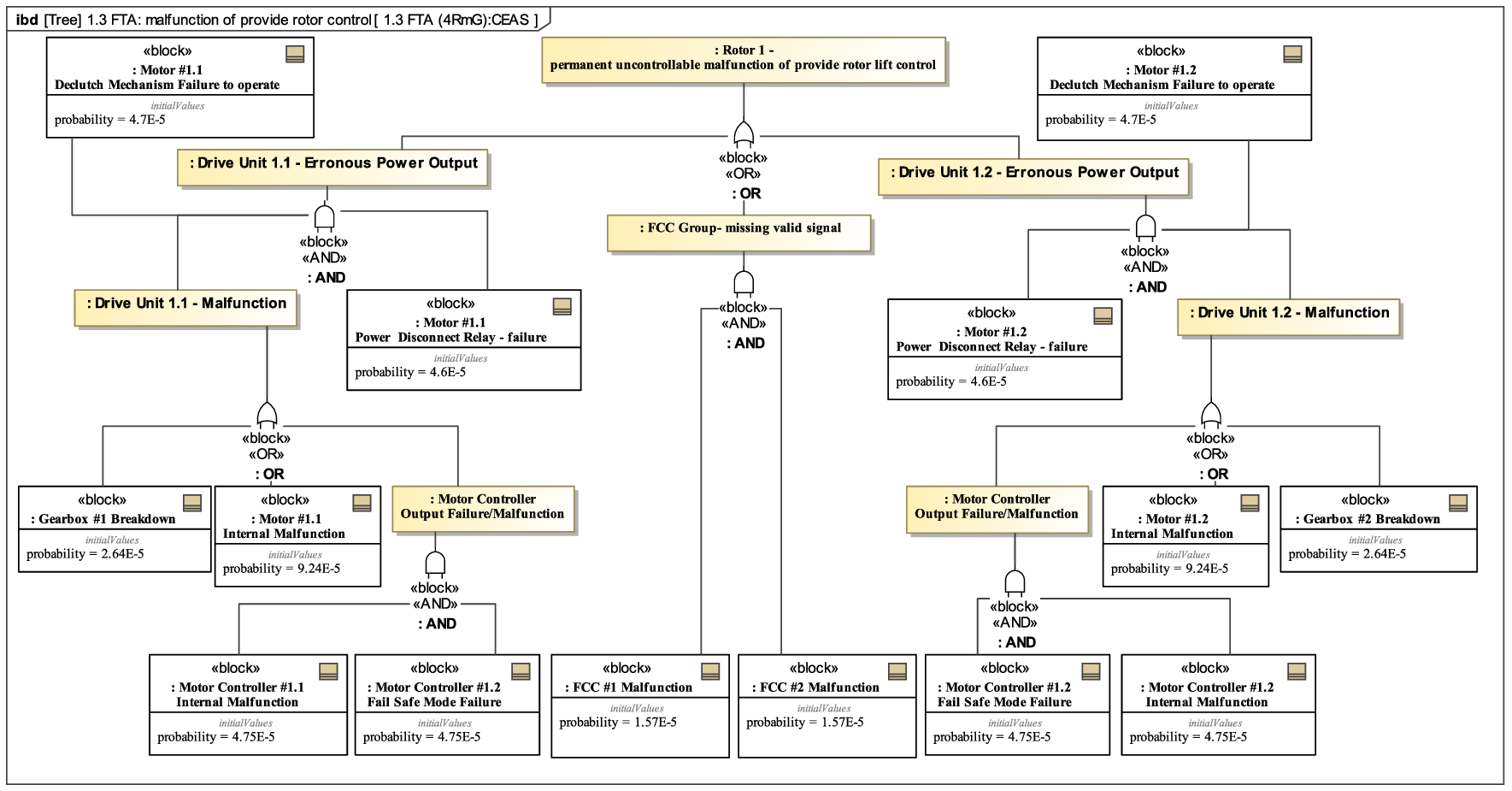}
    \caption{System FTA for the permanent incorrect operation of one rotor lift}
    \label{fig:SFTA}
\end{figure*}

For meeting the maximum allowable failure rate for the \textit{incorrect operation} of one rotor, it is essential to reduce the probability that an erroneous \acrshort{fcc} signal, motor controller output or motor output propagates up to the rotor. 
Firstly, the probability of an erroneous motor controller output is already reduced by the usage of a dual channel motor controller. Secondly, it must be prevented that any malfunction within the electric motor is passed to the rotor. Therefore, two masking strategies must be in place: once by disconnecting the power from the electric motor using an emergency power disconnect relay and secondly using a mechanical disconnect clutch that separates the motor output from the rotor shaft. Thirdly, the probability of an erroneous or missing valid \acrshort{fcc} command is reduced by implementing a triple modular redundant \acrshort{fcc} setup which enables determination of the correct output by majority voting. Each \acrshort{fcc} is then required to exhibit a malfunction or failure probability of less than $1.58\cdot10^{-5}$ per hour.\footnote{Based on the \acrshort{nprd}, this requirements is expected to be realistic as control board failure probabilities in average exhibit a failure rate of $1.2\cdot10^{-6}$ per hour \citep{Quanterion}, which therefore still allows for considering a deduction factor of 10 due to stresses that are caused by environmental influences and the operation.} By implementing these strategies both the probability for an erroneous power output of the driving units and the \acrshort{fcc} group is reduced below $10^{-10}$ per flight hour which reduces the \textit{incorrect operation} of one rotor to $2.46\cdot10^{-10}$ and therefore fulfils the allowable failure probability. The system \acrshort{fta} with the identified system components of the final propulsion system architecture contributing to a permanent \textit{incorrect operation} is shown in Fig. \ref{fig:SFTA}. 
The \textit{incorrect} or \textit{inadvertent operation} of the rear push propellers can be mitigated by the use of triple redundant FCC signals, a dual channel motor controller and a single disconnect option, like a disconnect relay. 

\begin{table}[htbp]
\caption{Extract of the PSSA results showing the expected system failure rates for the most limiting system level FTA top events} \label{tab:FailureRatesSFTA}
\begin{tabular}{@{}p{2.5cm} p{2cm} p{2cm}@{}}
\toprule
Functional Hazard & Max. allowable failure rate & Expected failure rate \\
\midrule
Loss of one rotor lift                              & $<2.5E-8$       &     $1,06 \cdot10^{-8}$ \\
Inadvertent ops. of one rotor         & $<1.0E-9$       &     $2.37~10^{-20}$  \\
Incorrect ops. of one rotor         & $<2.5E-10$      &     $2.46\cdot10^{-10}$ \\
Inadvertent ops. of one propeller   & $<5.0E-10$      &     $2.76\cdot10^{-15}$ \\
\botrule
\end{tabular}
\end{table}

\begin{table*}[htbp]
\caption{PASA results: expected failure rates for each identified catastrophic aircraft level function}\label{tab:3_FHAResults}%
\begin{tabular}{l p{2cm} p{2cm} p{2cm} p{2cm} p{2cm}}
\toprule
Function failure    & Provide lift   & Provide diff. thrust for pitch    & Provide diff. thrust for roll    & Provide diff. thrust for yaw    & Forward thrust  \\
\midrule
Total loss          & $4.49\cdot10^{-16}$    & $4.49\cdot10^{-16}$                      & $4.49\cdot10^{-16}$                      & hazardous                     & major         \\
Partial loss        & $4.24\cdot10^{-8}$     & $4.24\cdot10^{-8}$                         & $4.24\cdot10^{-8}$                         & n.a                           & n.a  \\
Incorrect ops.       & $9.86\cdot10^{-10}$  & $9.86\cdot10^{-10}$                      & $9.86\cdot10^{-10}$                      & hazardous                     & $5.52\cdot10^{-15}$         \\
Inadvertent ops.     & $9.86\cdot10^{-10}$    & $9.86\cdot10^{-10}$                     & $9.86\cdot10^{-10}$                      & hazardous                     & $5.52\cdot10^{-15}$          \\
Unable to stop      & $9.86\cdot10^{-10}$  & $9.86\cdot10^{-10}$                      & $9.86\cdot10^{-10}$                      & hazardous                     & $5.52\cdot10^{-15}$     \\
Unsym. partial loss & n.a.          & n.a.                               & n.a.                         & minor                         & minor        \\
Degradation         & major         & major                             & minor                             & minor                         & minor         \\
\botrule
\end{tabular}
\end{table*}

To also identify common causes of error, minimal cut sets were calculated and analysed within a reliability block diagram analysis. 
All derived requirements from the system level \acrshort{fta} and the minimal cut sets are listed below. Implementing these requirements for the overall system architecture ensures that the maximum allowable failure rates for the four system level \acrshort{fta} top events or respectively the basic events of the aircraft level \acrshort{fta} as shown in Table \ref{tab:FailureRatesSFTA} can be complied with.
\begin{enumerate}
    \item Each rotor requires a dual active redundant drive train. When a geared propulsion is chosen each drive train must also be equipped with a separate gearbox.
    \item Each rotor unit must be able to produce $\geq50\; \%$ of the total vehicle thrust required for hover for a prolonged time.
    \item For a short time interval, each rotor unit must be able to produce more than $50\; \%$ of the total hover thrust (ideal would be to provide $\geq50\; \%$ of the total vertical climb thrust) in order to break any vertical descent during landing. 
    \item Each motor unit must be able to be passivated and therefore be equipped with at least two means of decoupling, preferably a mechanical and electrical decoupling device. 
    \item Any internal fault of the motor control units must not lead to an unrecognized malfunction that propagates to the electric motor. Therefore, the motor control units should be designed as dual channel active passive units.
    \item The passive channel of the motor control unit acts as a fail-safe-backup-mode that activates in case of any loss of input signal from the \acrshort{fcc}s or in case of a motor control unit malfunction. In this state the motor control unit should command a constant motor rotational speed which corresponds to the hover state in normal flight. 
    \item Each motor control unit must be connected to at least two batteries or power supply busses to achieve a dual modular redundant power source. To prevent any common cause failures in total at least 4 battery packs are required for the four main rotors and one additional separate battery pack is required for the rear push power train. 
    \item One independent stand-alone battery source must be used to power both push-propeller units together.  
    \item Both rear propellers in combination must be able to create a vehicle yaw moment bigger than the resulting yaw moment of two shutdown concordant rotating main rotors.
    \item The \acrshort{fcc} setup must be triple modular redundant using majority voting. Each \acrshort{fcc} must exhibit a failure rate of $\leq1.58 \cdot10^{-5}$ per hour
\end{enumerate}

It becomes apparent that the loss of one rotor lift and the incorrect operation of one rotor providing lift are the most critical system design drivers for the propulsion system of the main rotors. As soon as the failure rate requirements are fulfilled for those events, the other functional hazards will be fulfilled as well. The push-propeller architecture however, is mainly driven by the functional hazard of an inadvertent operation of each propeller.

The resulting achievable failure rates for the top events of the aircraft level \acrshort{fta}, which were identified within the \acrshort{fha}, are summarized in Table \ref{tab:3_FHAResults}.
It is important to note that the architecture design and its failure probabilities are based on the following assumptions: 
\begin{enumerate}
    \item The loss of one main rotor does not lead to a catastrophic event as the opposite main rotor will be shut down to achieve equilibrium in pitch and roll. The resulting yaw moment is counteracted by the push propellers at the rear. The quadcopter therefore remains controllable and is able to continue safe flight and landing. 
    \item Each motor can be passivated by an own electric disconnect relay as well as a mechanical declutch mechanism. Thereby, it is assumed that for passivating the electric motor it is sufficient if either the mechanical declutch or the electric disconnect relay is activated. However, the control logic will only allow recovery of the electric motor drive to power the rotor if both the mechanical and electrical switches are closed. This is required in order to prevent a passivated malfunctioning electric motor drive to become operative due to an inadvertently closed electrical switch or mechanical clutch. 
\end{enumerate}

\FloatBarrier
\subsection{Propulsion System Component Sizing and Validation}\label{sec:SizingSimulation}
Within this section the main system groups of the propulsion system are sized, compared with off-the-shelf components and additionally simulated. At first the power and drive system is specified. Based on these results on the one hand the electrical system group with a primary focus on the batteries and on the other hand the thermal management system for the propulsion system are developed and analysed.
The simulation for each system group was carried out using the open modelling language Modelica within the Dymola environment by Dassault Syst\`{e}mes. 
\subsubsection{Sizing \& Simulation Power System}\label{sec:3.SizingPowerSystem}
As the power and drive system group consists of the motor controller, the electric motor, the gearbox and the rotor (see Fig. \ref{fig:LogicalArchitecture}), this section presents the specifications of these components for the main rotor and the rear push propeller drive system. The results are based on the following assumptions: 
\begin{enumerate}
    \item The main rotor drive is designed for a maximum rotor tip speed of $Ma_{tip}=0.45$ during normal operation and two rotor blades to ensure a quiet operation. The disc loading of 200 $N/m^2$ is chosen to aim for an energy efficient flight and minimise rotor losses. 
    \item The rear push propeller is designed for a cruise tip speed of $Ma_{tip}=0.5$, using a propeller with two blades and a propeller radius of 0.54 m to provide a cruise speed of 110 km/h at 3120 RPM.
    \item A system voltage for the power and drive system components is defined with 600 V.  
\end{enumerate}

Based on the safety assessment a total power of at least $200\; \%$ compared to the highest continuous required flight power needs to be provided by the four main rotor drives. For the quadcopter configuration this amounts to at least 450 kW that needs to be provided by eight electric motors. 

\begin{figure}[h!]
    \centering
    \includegraphics{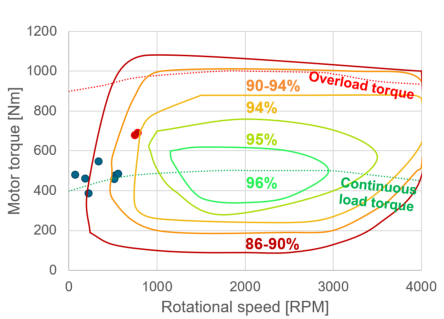}
    \caption{Representation of the electric motor operating points for a direct drive architecture (blue: hover, vertical climb, cruise climb, cruise, loiter, vertical descent; red: emergency hover and vertical climb) fitted within the efficiency map of a commercial off-the-shelf motor \cite{EMRAX.348} that is providing up to 1000 Nm torque}
    \label{fig:SizingDirectDrive}
\end{figure}

\begin{figure}[h!]
    \centering
    \includegraphics{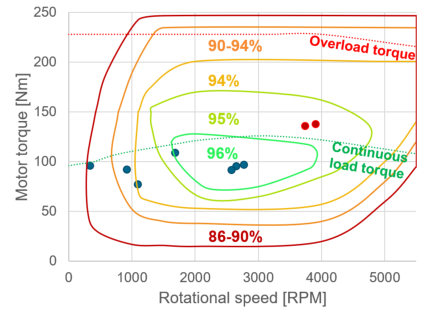}
    \caption{Representation of the electric motor operating points for a geared drive architecture with a 5:1 gear reduction ratio (blue: hover, vertical climb, cruise climb, cruise, loiter, vertical descent; red: emergency hover and vertical climb) fitted within the efficiency map of a commercial off-the-shelf motor \cite{EMRAX.228} that is providing up to $230\;Nm$ torque}
    \label{fig:SizingGearedDrive}
\end{figure}

During the sizing process a direct drive architecture was compared to a geared drive for the main rotor propulsion architecture. The comparison of a direct drive and a geared drive architecture assuming commercial off-the-shelf permanent magnet synchronous electric motors (PMSM) and planetary gearboxes\footnote{Based on an internal market study, suitable commercial off-the-shelf products were selected for this comparison.} clearly indicates the disadvantages of a direct drive. The currently available electric \acrshort{pmsm}s cannot be operated in their optimal efficiency range due to the low rotating speeds and high torque values that are required for the quadcopter main rotor propulsion. The efficiency of the direct drive lies between $ 85\;\% - 92\; \% $ as indicated in Fig. \ref{fig:SizingDirectDrive} whereas the geared drive is able to operate between $ 92\; \% - 96\; \%$ as indicated in Fig.\ref{fig:SizingGearedDrive}. Therefore, high thermal heat losses must be suspected using a direct drive.\footnote{Further information about the thermal management system are provided in the chapter \ref{sec:3.SizingTMSSystem}.} 

\begin{table*}[]
\caption{Recommended specifications for the power system components of the main rotor} \label{tab:SpecificationPowerSystem}
\noindent
\begin{tabularx}{1\textwidth}{@{}l X@{}} 
\toprule
Component & Specification  \\\hline
\midrule
\multicolumn{2}{@{} l}{Specifications of the main rotor power and drive system:}\\\midrule
Electric Motor  & \begin{mylist} \item Highest efficiency at 500-550 RPM (without gearbox) or 2650-2800 RPM (gearbox 5:1) with 100 Nm torque (hover \& vertical climb operating point)\end{mylist}\\ 
                & \begin{mylist} \item Continuous torque capability of $\geq$ 120 Nm and $\geq$ 145 Nm maximum peak torque\end{mylist}\\
                & \begin{mylist} \item Continuous power capability of 29 kW and a maximum peak power of 58 kW\end{mylist} \\
                & \begin{mylist} \item Max RPM at $\geq$ 780 RPM (without gearbox) or 3905 RPM (with gearbox 5:1)\end{mylist}\\ 
Gearbox         & \begin{mylist} \item Reduction gear ratio of 5:1\end{mylist} \\
                & \begin{mylist} \item Input rotating speed range of 330-2770 RPM (normal ops), up to 3905 RPM in irregular operation \end{mylist}\\
                & \begin{mylist} \item Output rotating speed range 66-554 RPM (normal ops), up to 781 RPM in irregular operation\end{mylist}\\
                & \begin{mylist} \item Equivalent output torque of $\geq$ 455 Nm\end{mylist}\\
                & \begin{mylist} \item Maximum peak output torque of $\geq$ 700 Nm\end{mylist}\\ 
Motor Controller & \begin{mylist}\item Continuous power of $\geq$ 30 kW\end{mylist}\\
                & \begin{mylist} \item Maximum peak power of $\geq$61 kW\end{mylist}\\ \midrule
\multicolumn{2}{@{} l}{Specifications of the push propeller power and drive system:}\\\midrule
Electric Motor  & \begin{mylist} \item Highest efficiency at 3120 RPM with 99 Nm torque (cruise operating point)\end{mylist}\\
                & \begin{mylist} \item Continuous torque capability of 99 Nm and a maximum peak torque of 125 Nm\end{mylist}\\
                & \begin{mylist} \item Continuous power capability of 32 kW and a maximum power of 60 kW\end{mylist}\\
                & \begin{mylist} \item Max RPM of $\geq$ 4576 RPM\end{mylist}\\ 
Motor Controller& \begin{mylist} \item Continuous power of $\geq$ 34 kW and maximum peak power of $\geq$ 63 kW\end{mylist}\\
                & \begin{mylist} \item Continuous motor current $\geq$ 90 A and maximum motor current $\geq$ 108 A\end{mylist}\\

\botrule
\end{tabularx}
\end{table*}

\begin{figure}[h!]
    \centering
    \includegraphics[width=\linewidth]{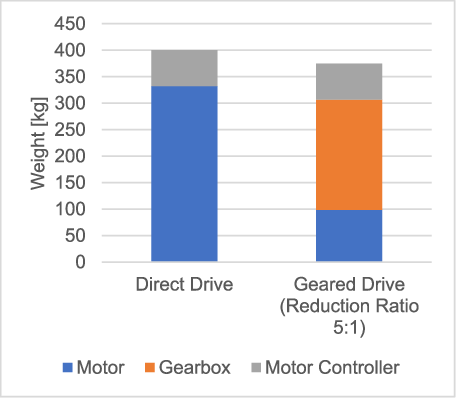}
    \caption{Weight comparison of a direct drive and geared propulsion system for the main rotor}
    \label{fig:WeightComparison}
\end{figure}
In terms of the main propulsion system weight (consisting of the electric motor, motor controller and gearbox) the geared propulsion architecture exhibits a weight of 374 kg compared to 400 kg for the direct drive. As it can be seen in Fig. \ref{fig:WeightComparison} the weight of the direct drive powertrain is mainly driven by the electric motor. While the electric motor for the geared powertrain is approximately $70\; \%$ lighter, heavy gearboxes are required to provide sufficient torque capability. Still, a weight saving of almost $7\; \%$ is estimated using the geared drive.

This comparison of the motor efficiency and the expected propulsion system weight shows that the usage of a geared drive is advisable to save weight and thermal losses. The gearbox as an additional component of the powertrain therefore needs to be considered in the safety and reliability analysis. 

The propulsion system for the rear push propulsors is sized to provide the highest efficiency during the cruise flight. This requires an electric motor to provide the highest efficiency at 3120 RPM and 99 Nm torque. In total the electric motors for the rear push drive are expected to weigh 42 kg and amount to $58\; \%$ of the rear propulsion weight. As the rear drive system does not require a gearbox, the motor controllers make up for the remaining $42\; \%$.

The specifications for each component of the main rotor and rear propeller power and drive system are listed in Table \ref{tab:SpecificationPowerSystem}.\\

The simulation of the power and drive system has shown that the powertrain components with the above-mentioned specifications are suitably sized for powering the main rotors as well as the push propellers. Also, in failure conditions the main rotors can still be accelerated to the required rotational speeds. The preliminary dynamic simulation of the rotor rotational speed following control signals shows rise times in the magnitude of tenth of seconds, depending on the step size. However, whether or not this rise time of the rotor is sufficient to achieve quick response times of the total vehicle and therefore to achieve good handling qualities is still under further analysis\footnote{For further information concerning the controllability and handling quality of RPM controlled rotors within a quadcopter it is referred to \cite{Atci.2022}.}. 

\vspace{5mm} 
Based on these analyses the following implications can be drawn for the propulsion system architecture: 
\begin{itemize}
\item Using a gearbox is recommended for the main rotor drive system. 
\end{itemize}

\FloatBarrier
\subsubsection{Sizing \& Simulation Electrical System} \label{sec:3.SizingElectricalSystem}

\begin{figure*}[htpb]
    \centering
    \includegraphics{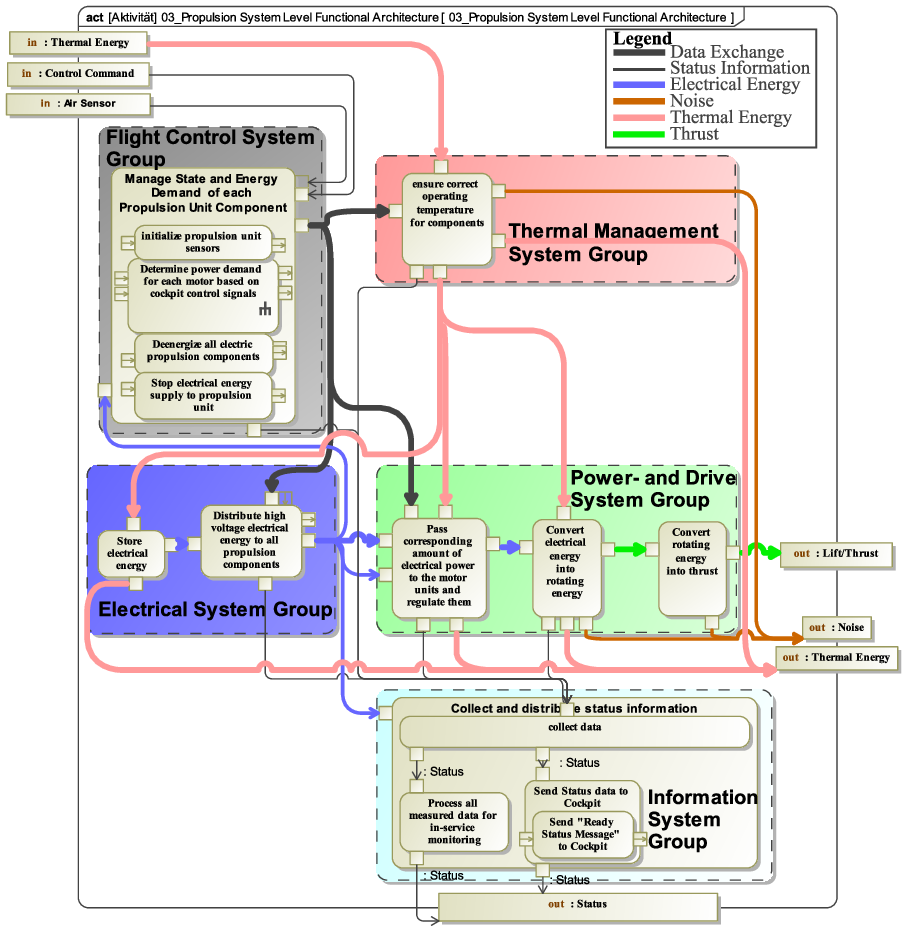}
    \caption{Functional block diagram of the propulsion system}
    \label{fig:FunctionalArchitecture}
\end{figure*}

Based on the functional propulsion architecture of Fig. \ref{fig:FunctionalArchitecture} and the correspondingly derived logical architecture of  Fig. \ref{fig:LogicalArchitecture} a battery storage system is required to power the propulsion system. As described by the requirement no. 7 of the system and reliability analysis of section \ref{sec:3.Safety} in total four identical batteries are required for powering the main rotors and one common battery pack is required for the rear push-propeller drive trains. 
Within this section an initial sizing of the energy storage system is conducted. This includes an analysis of the required energy, the battery pack size and weight by considering failure conditions of the propulsion system during flight. 

Looking into the electrical power distribution of the main rotors, each motor controller needs to be able to receive power from an alternate battery source in case its main battery source is unavailable, in order to fulfil the requirement of section \ref{sec:3.Safety}. Therefore, an allocation as shown in Table \ref{tab:BatteryDistribution} is chosen. The Table indicates that for the rotor drive number 1 the motor controller 1.1. receives mainly power from battery 1 but can be switched to battery 3, whereas motor controller 1.2 is powered by battery 2 and 4. Based on this allocation each battery continuously powers two motor controllers, with each motor controller consuming a maximum of 30 kW during normal operations. 
\begin{table}[htpb]
\caption{Allocation of the four main battery packs to the motor controllers of each main rotor drive system} \label{tab:BatteryDistribution}
\begin{tabular}{@{}lllll@{}}
\toprule
  & Rotor 1 & Rotor 2 & Rotor 3 & Rotor 4 \\
\midrule
MC x.1              & BAT 1 & BAT 1 & BAT 3 & BAT 2\\ 
MC x.1 ALT          & BAT 3 & BAT 2 & BAT 2 & BAT 3\\
MC x.2              & BAT 2 & BAT 3 & BAT 4 & BAT 4\\
MC x.2 ALT          & BAT 4 & BAT 4 & BAT 1 & BAT 1\\
\botrule
\end{tabular}
\end{table}

\begin{figure}[htpb]
    \centering
    \includegraphics{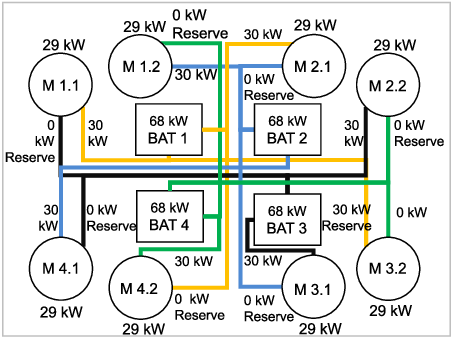}
    \caption{Power allocation of the main rotor battery packs during normal operation. Above or below each motor symbol the maximum required power of each electric motor is indicated for the shown operating state. The required motor controller power is shown next to the corresponding power lines of the motor symbol. On top of each battery pack the maximum available power of each battery pack is listed.}
    \label{fig:BatteryAllocationNormalOps}
\end{figure}

Fig. \ref{fig:BatteryAllocationNormalOps} shows the power allocation of the main rotor battery packs during normal operation. 
Based on the allocation, the sizing of the battery packs can be conducted which is influenced by the following requirements: 
\begin{enumerate}
\item The batteries must provide the total energy required for fulfilling the flight mission during normal ops. 
\item The capacity of each battery pack must be sufficient to provide sufficient power for the connected motor controllers during normal ops. 
\item	The battery packs must provide enough energy and power for enabling a continued safe flight and landing during any failure condition
\end{enumerate}
Each requirement is now analysed separately. As a basis the Panasonic 18650 battery cells are used \cite{Panasonic.2013}. Since a system voltage $U_{sys}$ of 600 V is chosen, each battery pack should provide 600 V which requires 167 cells connected in series. 
\vspace{5mm}

\textbf{Battery sizing based on the required total energy and power for normal operation:}\\
Based on the energy requirement of 19.7 kWh\footnote{The energy requirement was derived from the initial sizing process and validated with the developed simulation model} for powering two drive units during normal operation for the total flight mission $E_{BS}$, the required battery pack capacity $C_{BP,E}$ amounts to 32.8 Ah.  
\[
C_{BP,E}=\frac{E_{BP}}{U_{sys}} = \frac{19.7\; kWh}{600\; V}=32.8\; Ah
\]
The battery capacity based on the two supplied drive units with a required maximum power $P_{max}$ of 30 kW each for normal operations amounts to 29.3 Ah.
\[
C_{BP,P}=\frac{P_{max}}{\xi\cdot U_{sys}} = \frac{2\cdot\ 30\; kW}{3.448\frac{1}{h}\cdot 600\; V}=29.3\; Ah 
\]
using the cell discharge rate $\xi$ that is composed of the Panasonic rated battery cell current $i_{Batt}$ and rated capacitance $c_{Nenn}$
\[
\xi=\frac{i_{Batt}}{c_{Nenn}}=\frac{10\; A}{2.9\; Ah}=3.448 \frac{1}{h}.
\]
Sizing the battery with the higher of both capacity requirements of 32.8 Ah results in a battery pack that is able to provide 68 kW discharge power. 

  \begin{align*} 
     P_{available}  & = C_{BP}\cdot\xi\cdot U_{Sys}\\
                    & = 32.8\; Ah \cdot 3.448\frac{1}{h}\cdot\; 600V\\
                    & = 68\; kW.
  \end{align*}

\begin{figure}[htpb]
\centering
    \begin{subfigure}{0.45\textwidth}
    \centering
        \includegraphics[width=\linewidth]{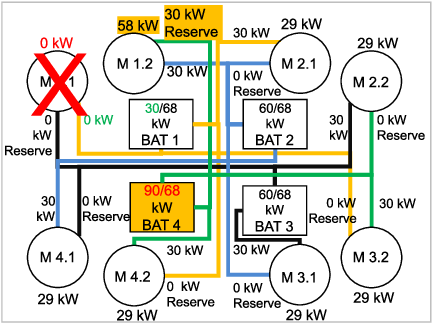}
        \caption{Failure case: single rotor drive unit loss caused by a motor or motor controller failure (not optimized)}
        \label{fig:BatteryAllocationMotorFailure}
    \end{subfigure}

    \begin{subfigure}{0.45\textwidth}
        \centering
        \includegraphics[width=\linewidth]{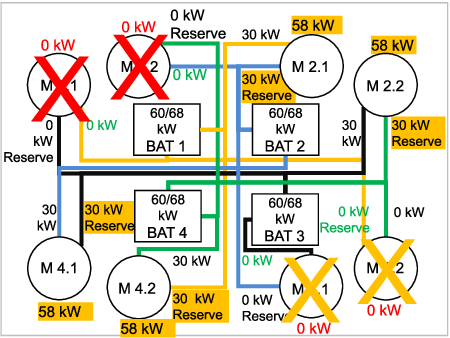}
        \caption{Failure case: complete main rotor drive loss}
        \label{fig:BatteryAllocationRotorFailure}
    \end{subfigure}

    \begin{subfigure}{0.45\textwidth}
        \centering
        \includegraphics[width=\linewidth]{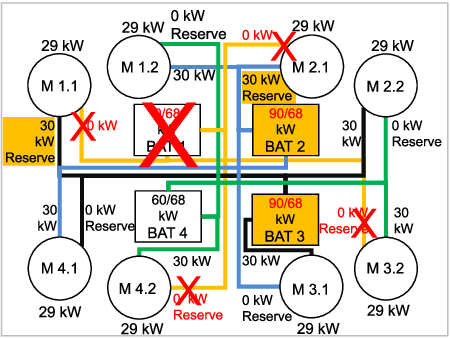}
        \caption{Failure case: single battery pack loss}
        \label{fig:BatteryAllocationBatFailure}
    \end{subfigure}
\caption{Power distribution of the main rotor battery packs for different failure cases}
\label{fig:BatteryAllocationFailures}
\end{figure}

\textbf{Battery sizing based on the energy and power requirement during failure conditions.} \\
The three most restrictive failure conditions are analysed and their effects on the power distribution of the battery packs shown in figure \ref{fig:BatteryAllocationFailures}: 
\begin{itemize}
\item Failure of one rotor drive unit due to motor or motor controller unit failure
\item Failure of one rotor unit
\item Failure of one battery pack 
\end{itemize}
\vspace{4mm}

\textbf{Power requirement:}
During a single motor unit loss or a failure of one battery pack at least one of the remaining battery packs is required to supply at least 90 kW power as indicated in Fig.  \ref{fig:BatteryAllocationMotorFailure} and \ref{fig:BatteryAllocationBatFailure}. 
In order to provide 90 kW battery pack power, a pack capacity of 43.5 Ah is required. 
\[
C_{BS,P}=\frac{P_{min}}{\xi\cdot U_{sys}} = \frac{90\; kW}{3.448\frac{1}{h}\cdot 600\; V}=43.5\; Ah 
\]

During a single rotor loss, however, the available 68 kW pack power are sufficient to cope with this failure. 
\vspace{5mm}

\textbf{Energy requirement:}
Whether or not the battery capacity is sufficient for reaching a suitable airfield even during failure conditions depends on the type of failure condition, the time at which the failure occurs and the intended emergency flight procedure. Herein the most unfavourable conditions are assumed: 
\begin{itemize}
\item Based on the previous failure analysis the type of failure condition is assumed to draw 90 kW power of a single battery pack for the remainder of the flight.
\item The failure condition occurs on the third flight of the total flight mission at the equal time point (ETP)\footnote{The ETP defines the point within each flight, where the time to reach the next suitable airfield equals the time to return to last overflown or departed airfield.}. The ETP for the defined flight is reached 5.7 min after start of the third flight segment. Up to this point approximately 12.2 kWh are already consumed, which includes the energy for flight one and two and the energy up to the ETP. \footnote{Assuming the flights one and two were carried out as planned and no contingency energy has been used.}  
\item There is no other closer landing site available than the intended destination. Therefore, the flight is continued to the destination. 
\end{itemize}
With these assumptions another 8.5 kWh are required, which is the equivalent energy to reach the destination airfield within 5.7 minutes. Therefore, a battery capacity of at least 34.3 Ah or 20.6 kWh are required. All battery pack capacity requirements of the normal and failure condition operation are summed up in Table \ref{tab:BatteryCapacitySummary}.

\begin{table}[]
\caption{Summarized capacity requirements for each main rotor battery pack based on the energy and power requirements} \label{tab:BatteryCapacitySummary}
\begin{tabular}{@{}lll@{}}
\toprule
Capacity Requirement   & Battery capacity & Destination \\
based on            &                  & reachable?\footnotemark[1]\\
\midrule
Energy required & 32.8 Ah & no \\
(normal ops) \\ 
Energy required & 34.3 Ah & yes\footnotemark[2]\\ 
(emergency ops)\\
Power required & 29.3 Ah & no \\ 
(normal ops) \\
Power required & 43.5 Ah & yes \\ 
(emergency ops)\\
\botrule
\end{tabular}
\footnotetext[1]{Indicates whether destination can be reached in case of emergency operation starting within the third flight segment.}
\footnotetext[2]{Defines the absolute minimum}
\end{table}

It becomes apparent that a main rotor battery pack capacity of 43.5 Ah is required and driven by the maximum power requirement during the failure condition caused by a single rotor drive unit loss or a battery pack loss. Under normal operating conditions a pack with this capacity will be discharged down to $25\;\%$ after having completed the flight mission and used 20 min reserve flight time and will have 9.2 min of flight time available at the occurrence of a failure condition at the ETP during the third flight.
A battery with these specifications can be composed of 15 of the Panasonic cells connected in parallel and 167 in series, so that in total 2505 cells\footnote{The effect of battery degradation has so far not been in the scope of this research.} are used. Each battery pack then weighs 120 kg. 
\vspace{5mm}

\textbf{Battery sizing for the rear push propulsion system:}\\
The same assessment is conducted for the push propeller propulsion battery pack, which must be able to provide at least 68 kW power during normal operation and 126 kW power during an emergency condition at which a main rotor has failed. The results are summarized in Table \ref{tab:BatteryCapacitySummaryPush}. 

\begin{table}[]
\caption{Summarized capacity requirements for the push propulsion battery pack based on the energy and power requirements} \label{tab:BatteryCapacitySummaryPush}
\begin{tabular}{@{}lll@{}}
\toprule
Capacity Requirement   & Battery capacity & Destination \\
based on            &                  & reachable?\footnotemark[1]\\
\midrule
Energy required & 70.0 Ah & no \\ 
(normal ops) \\
Energy required & 88.8 Ah & yes \footnotemark[2]\\ 
(emergency ops)\\ 
Power required & 32.6 Ah & no \\ 
(normal ops) \\
Power required & 61.1 Ah & no \\
(emergency ops) \\
\botrule
\end{tabular}
\footnotetext[1]{Indicates whether destination can be reached in case of emergency operation starting within the third flight segment.}
\footnotetext[2]{Defines the absolute minimum}
\end{table}
The battery pack capacity in this case is mainly driven by the energy requirement during the failure condition at the most unfavourable point of time. Therefore, the battery is sized with a capacity of at least 88.8 Ah\footnote{As the battery sizing based on the normal operation results in more than $20\;\%$ less required capacity than based on the highest failure case capacity requirement, no additional reserve of $20\;\%$ for the battery capacity is included} respectively 54.1 kWh, which results in having 31 parallel cells and 167 cells connected in series. In total 5177 cells are required, which results in a capacity of 90 Ah, with 186 kW available power. 
During normal operation the battery pack is discharged down to $37\;\%$ in case the 20 minutes loiter time had been utilized during flight. At the ETP the battery provides enough energy to power the push propulsion system for another 6 minutes in case of emergency operations, which is sufficient to reach the landing site.  

Further optimization of the battery pack size could be gained by an intelligent battery management system that allows an interconnection between all battery packs and, thereby, allocates energy and power requirements smartly between all battery packs. 

The simulation of the battery packs in combination with the previous power and drive system has indicated that the battery is sufficiently sized to provide energy for the whole flight mission.
The heat development within each battery pack is analysed within the next section and the requirements for a cooling system are derived,

\vspace{5mm} 
Summary of the derived specifications for the propulsion architecture: 
\begin{itemize}
\item{Each main rotor battery pack requires a battery capacity of at least 43.5 Ah and should be able to provide at least 90 kW power.}
\item{The push-propeller battery pack requires a battery capacity of at least 88.8 Ah and should be able to provide at least 68 kW power.}
\end{itemize}
\FloatBarrier
\subsubsection{Sizing \& Simulation Thermal Management System}\label{sec:3.SizingTMSSystem}
Within this section the results of the heat development simulation within the power and drive system components are presented and initial system requirements for a thermal management system derived. As a result from previous section, this analysis is based on the propulsion design which incorporates a geared drive propulsion system. 
Whereas the gearbox, as an encapsulated system, can be expected to be self-cooling and lubricating, the electric motor and motor controller are expected to be combinable within a cooling system due to their similar requirements. For the battery packs, however, a separate thermal management system is expected to be required, which allows for heating at low temperatures and cooling during operation at high operating temperatures, due to the narrow optimal battery operating temperature range of 20-40 °C. When assuming \acrshort{vtol} operation at the warmest areas within Europe, ambient temperatures of up to 42.7 °C\footnote{The highest measured temperature of the European city Madrid within the years 2017 and 2022 \cite{Meteostat}.} are taken into account during the design process of the propulsion and cooling system.

\begin{figure}[htpb]
    \centering
    \includegraphics{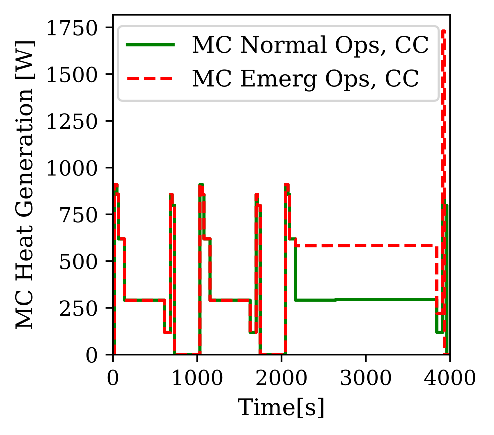}
    \caption{Heat generation of the electric motor during normal and emergency operation which is assumed to start at the end of the vertical flight phase of the third flight.}
    \label{fig:HeatGenerationMotor}
\end{figure}

\begin{figure}[htpb]
    \centering
    \includegraphics{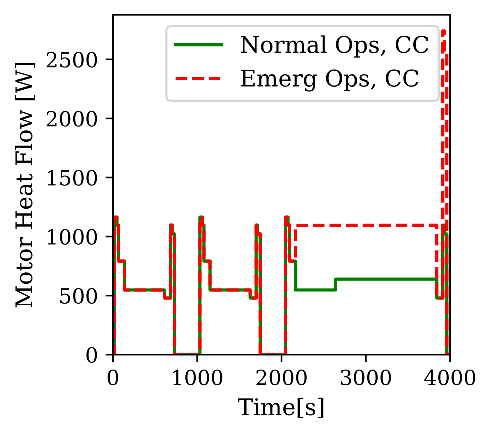}
    \caption{Heat generation of the motor controller during normal and emergency operation which is assumed to start at the end of the vertical flight phase of the third flight.}
    \label{fig:HeatGenerationMC}
\end{figure}

Initially the thermal management system requirements for the motor and motor controller of one rotor drive system are evaluated. Analysing the amount of generated heat within the electric motor and motor controller gives the following results. During the phase of the highest power requirement, the vertical climb phase, the motor controller and the electric motor are expected to produce up to 1.1 kW and 1.4 kW heat respectively under normal operating conditions as shown in Fig. \ref{fig:HeatGenerationMotor} and Fig. \ref{fig:HeatGenerationMC}. During emergency conditions, in which only one electric motor is left driving a main rotor, the heat output amounts to 1.8 kW and 2.9 kW respectively. 

\begin{figure*}[htpb]
    \centering
    \includegraphics{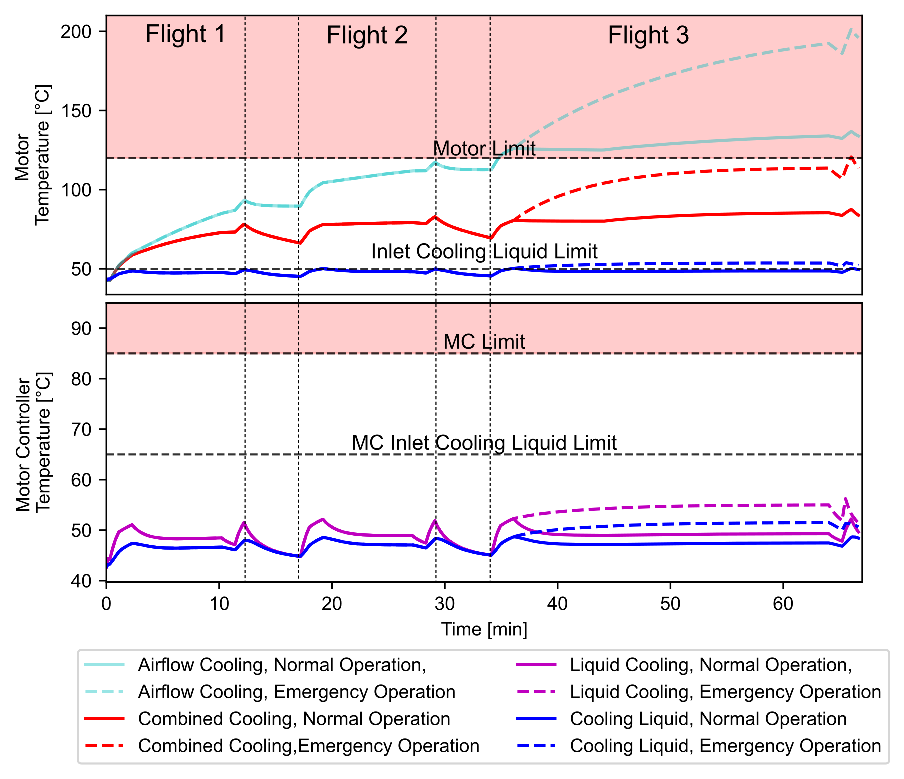}
    \caption{Temperature development within the electric motor copper windings using only air flow cooling (AF) or a combined cooling (CC) compared to the temperature development of the motor controller using liquid cooling.}
    \label{fig:M+MCTempDevelopment}
\end{figure*}

As the intended electric motor provides means for airflow cooling, its effect on the heat development within the electric motor components is initially analysed. With an ambient air flow that is based on the flight mission, the flight speeds, the vehicle rotor configuration and the cooling tubes geometry, the temperature within the copper windings of the electric motor will still rise over 120 °C during the transition of hover to cruise climb of the third flight within the flight mission after 2088 seconds and reach a maximum of 136 °C as shown in Fig. \ref{fig:M+MCTempDevelopment}. 
Thus, the temperatures within the electric motor cannot be kept below 120°C during normal operation using only the ambient air flow. Operating in the emergency rating, in which one electric motor must provide the full power for a single rotor, the temperatures within the electric motor would even rise over 200 °C during the third flight of the mission (see Fig. \ref{fig:M+MCTempDevelopment}). Consequently, the electric motor cooling system must be complemented by an additional liquid cooling system. As the analysed motor controller also requires liquid cooling according to the manufacturer's data sheet in order to keep its temperature below 85 °C, the motor controllers and corresponding electric motors of one rotor drive unit are combined within the same liquid cooling system. As the four rotor drive units will be located below each rotor and therefore be located distant to each other, a separate cooling system for each rotor drive unit is recommended.  

\begin{figure}[htpb]
    \centering
    \includegraphics{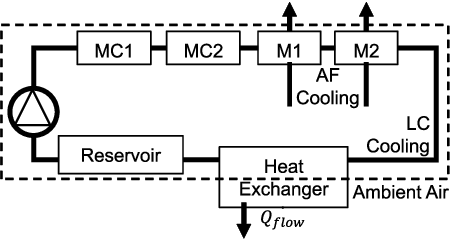}
    \caption{Schematic view of the combined cooling system for the electric motor and motor controller consisting of a liquid cooling cycle and airflow cooling.}
    \label{fig:CoolingTopologie}
\end{figure}

As a conclusion the following requirements must be fulfilled by a combined ambient air flow and liquid cooling system for the off-the-shelf analysed electric motor and motor controllers. 
\begin{enumerate}
    \item The electric motor operating temperature must be kept below 120 °C also during emergency conditions while the motor is operating in emergency rating.
    \item The motor controller operating temperature must be kept below 85 °C during all operating conditions (normal and emergency rating).
    \item As the motor controller can be operated with a liquid cooling temperature of a maximum of 65 °C and the electric motor requires liquid cooling temperatures below 50 °C, the thermal management system needs to keep the cooling liquid temperature below 65 °C when passing the motor controller and 50 °C when passing the electric motors.
    \item The maximum volume flow of cooling liquid for the motor controller is 6-12 l/min whereas the electric motor can only withstand 6-8 l/min.
    \item The maximum input pressure of the liquid cooling shall not exceed 2 bar when entering each electric motor.
\end{enumerate}
\vspace{5mm} 
One exemplary cooling topology that is able to fulfil these requirements consists of the motor controllers and electric motors of each rotor being connected in series as shown in Fig. \ref{fig:CoolingTopologie}. As the motor controllers can withstand higher liquid input pressures and emit less heat they are placed at the beginning of the cooling flow. The electric motors can then be placed downstream.\footnote{A possible alternative would be to place the electric motors in parallel downstream.}

\begin{figure*}[htpb]
    \centering
    \includegraphics{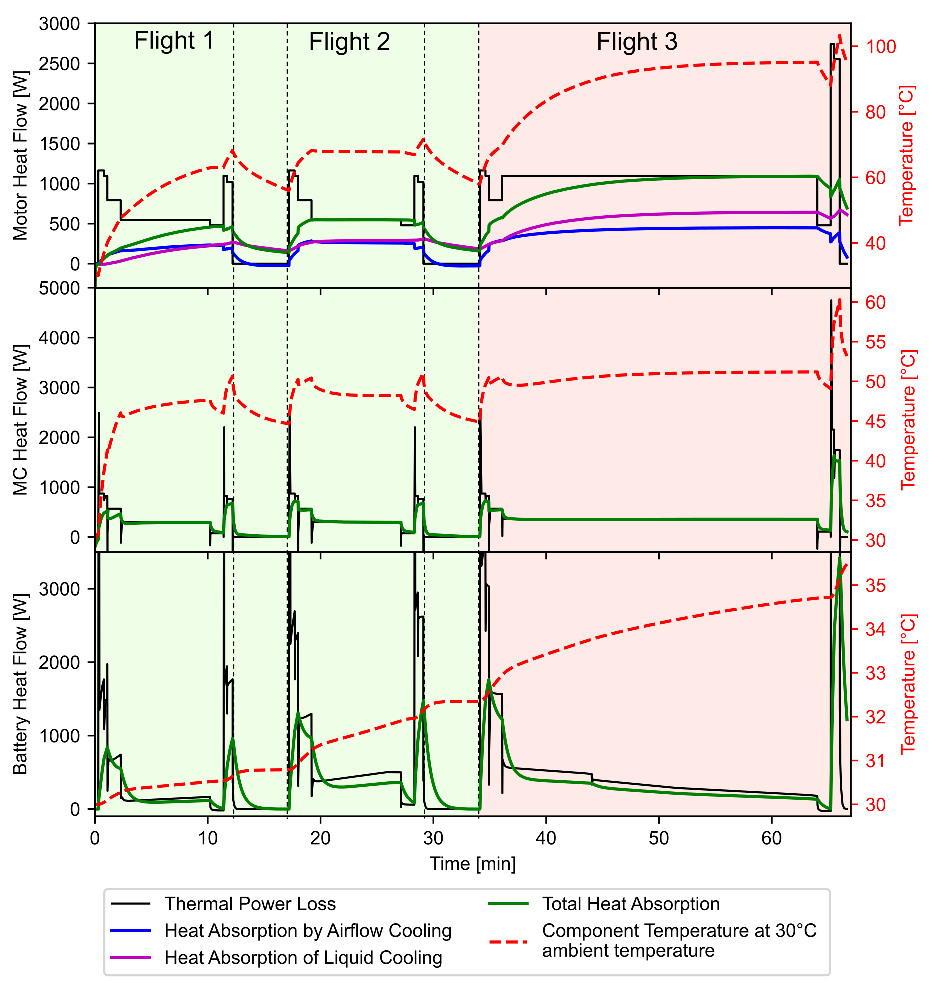}
    \caption{Heat power loss within the electric motor, motor controller and battery pack and its dissipation via ambient air flow and / or liquid cooling at an ambient temperature of 30 °C}
    \label{fig:HeatFlowM+MC+Bat}
\end{figure*}

The combined cooling system (CC) consisting of airflow cooling (AF) and liquid cooling (LC) is designed with the following specifications:
\begin{itemize}
    \item Cooling fluid: Glysantin G40
    \item Cooling liquid flow: 0.14 kg/s
    \item Cooling air flow: dependent on flight phase - maximum available 2.3 $m^3/kg$
    \item Heat exchanger size: $0.3\cdot0.3\cdot0.3$ m
    \item Heat exchanger weight: 3.8 kg
\end{itemize}
\vspace{5mm} 
The temperature development within the electric motor using this cooling topology of combined cooling for the electric motor and liquid cooling for the motor controller are shown in Fig. \ref{fig:M+MCTempDevelopment}. It can be seen, that the temperature within the electric motor stays below 120 °C during all operating conditions, even in emergency conditions at the most unfavourable situation during the flight mission when using a combined liquid and air-cooling system. Fig. \ref{fig:HeatFlowM+MC+Bat} indicates how the generated heat within the electric motor is absorbed by the ambient air flow and the liquid flow. As not all of the generated heat within the first flight can be dissipated, the temperature within the electric motor components rises during the flight mission.
The inlet liquid cooling temperature for the electric motor as shown in Fig. \ref{fig:M+MCTempDevelopment}, however, almost reaches the maximum manufacturer's recommendation of 50 °C during normal operation. Operating in emergency rating the inlet temperature even rises up to 54 °C and therefore exceeds the manufacturer's limit of 50 °C.

The temperature within the motor controller stays well below its limit of 85 °C as shown in Fig. \ref{fig:M+MCTempDevelopment}. All of the heat is absorbed by the liquid cooling flow. During the whole flight mission, the maximum inlet temperature of the liquid fluid for the second downstream motor controller (MC2) reaches a maximum of 60 °C (during emergency ops) and therefore stays below the required 65 °C.

To prevent an excessive heat build-up during the ground phases of the flight mission, in which the vehicle is not moving and therefore receives no cooling air flow, it is essential to keep up the liquid cooling flow as well as the cooling air flow. Therefore, the liquid cooling pump needs to be operative. Additionally, a ground fan, should be installed to facilitate the heat transfer within the heat exchanger. Using no ground cooling the temperature of all components within the cooling system would heat up to over 65 °C during the five minutes ground phase. The same effect can be observed after termination of the flight mission, where a temperature of 77 °C can be observed across all components after 30 minutes. By increasing the ground cooling time up to 30 minutes the overall temperature can be kept below 50 °C.  
\vspace{5mm} 

\begin{figure}[htpb]
    \centering
    \includegraphics{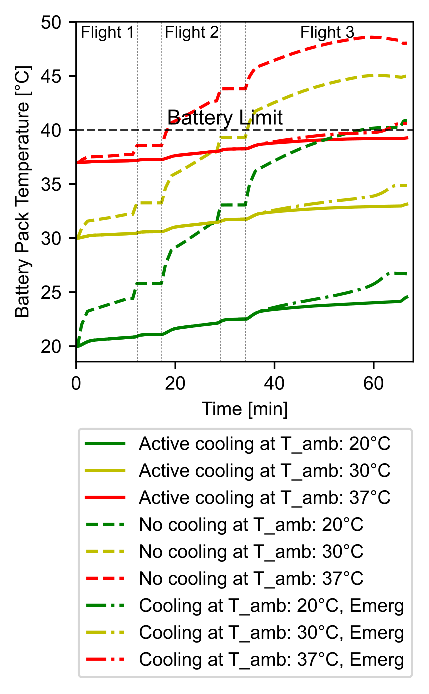}
    \caption{Temperature development within each battery pack at different ambient temperatures with and without cooling during normal operation}
    \label{fig:BatteryTemp}
\end{figure}

In the following, the thermal management system requirements for the battery packs are evaluated. Fig. \ref{fig:BatteryTemp} compares the temperature development within each battery pack during normal operation with and without any cooling system. 
It becomes clearly visible that the batteries cannot be operated without a cooling system as the battery pack temperature will exceed the 40 °C temperatures limit even at 20 °C ambient temperatures. Using a liquid cooling circuit with a glycol-water mixture of 20:80 that is channelled along each battery cell, the temperatures of each battery pack can be kept below 40 °C during normal operation if the ambient temperature does not rise above 37 °C.
However, under emergency operation, in which one battery pack has failed during the transition to cruise on the third flight segment, the battery pack temperature of two battery packs (refer to Fig. \ref{fig:BatteryAllocationBatFailure}) will even increase up to 40.6 °C at ambient temperatures of 37 °C. The ambient temperature has to stay below 36.2 °C in order to ensure a battery pack temperature below 40 °C during emergency conditions using a liquid cooling circuit. The liquid cooling circuit used within this simulation has the following properties: 
\begin{itemize}
\item Outer cooling circuit - ambient air volume flow: 0.3 $m^3/s$ 
\item Inner cooling circuit - liquid volume flow: $6.8\cdot 10^{-5}\; m^3/s$ 
\item Inner cooling fluid: glycol-water mixture 20:80  
\end{itemize}
For higher ambient temperatures a refrigeration cycle is required which is part of future research. 
\vspace{5mm} 

A summarizing overview about the expected behaviour of the electric motors, motor controllers and main rotor battery packs as well as their thermal management systems during normal and emergency operation is given in Fig. \ref{fig:HeatFlowM+MC+Bat}. Besides the expected heat flow also the amount of heat absorption of each used cooling method is shown. Additionally, the temperature development of each component is indicated for a typical hot summer day with 30 °C ambient temperature. 
\vspace{5mm} 

Based on the analyses of this section the following implications can be drawn for the propulsion system architecture: 
\begin{enumerate}
\item Each propulsor requires a separate cooling system 
\item For each electric motor, that was chosen and analysed herein, a combination of liquid cooling and air cooling is required as only air cooling is not sufficient to cool the electric motor sufficiently.
\item Each motor controller requires a liquid cooling system
\item The motor controller and electric motor can be cooled using the same liquid cooling system.
\item Each liquid cooling system consists of the components: pump, cooling fluid reservoir, heat exchanger.
\item The cooling system should stay operative after each flight on ground for up to 30 minutes to prevent the cooling liquid to exceed the motor inlet temperature of 50 °C and to absorb the stored thermal energy within the electric motor and motor controller. 
\item The battery needs an own thermal management system that is capable of cooling and heating. If operating at ambient temperatures of 20 - 36.2 °C each battery pack can be cooled using a liquid cooling cycle. For ambient temperatures outside this range the thermal management system still needs to be analysed and designed.
\end{enumerate}
\FloatBarrier
\subsection{Final Architecture}\label{sec:3.FinalArchitecture}
\begin{figure*}[htpb]
    \centering
    \includegraphics{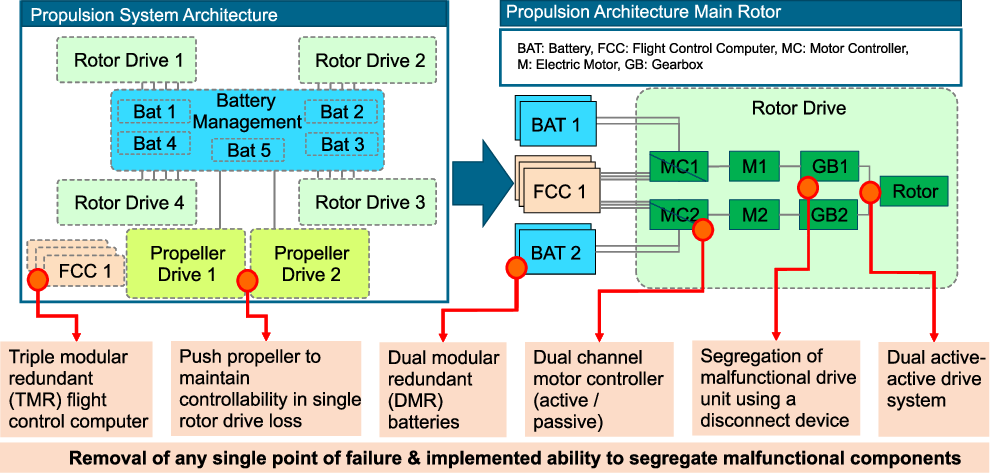}
    \caption{Schematic representation of the overall quadcopter propulsion architecture and its implemented safety measures, excluding the thermal management system.}
    \label{fig:FinalArchitecture}
\end{figure*}
This section presents how the requirements of the previous sections were implemented in the final propulsion system design. This architecture is expected to fulfil the safety reliability requirements. 
A schematic representation of the propulsion system architecture for the quadcopter is shown in Fig. \ref{fig:FinalArchitecture}. On the left side the main propulsion systems are shown with the four main rotor drives that are supplemented by two push drives. Each drive unit is connected with the battery units. In total at least three \acrshort{fcc} are provided. On the right the propulsion architecture for each main rotor is depicted in more detail. It shows that the main rotor propulsion architecture is composed of two electric motors that drive the rotor through two separate gearboxes. Each motor is driven by its own motor controller, while each motor controller is backed up with a passive control board that takes over, in case faulty signals are sent by the primary controller or even the connection is lost. Each motor controller receives power from one of the four main batteries and can be switched to an alternate battery source if necessary. Additionally, each motor controller receives inputs from all three \acrshort{fcc} and determines the valid \acrshort{fcc} input by majority voting. 
The main criticality of the propulsion system of the push-propeller drive is the \textit{incorrect operation} including the \textit{inadvertent} or \textit{unable to stop operation} which are classified as a potentially catastrophic event. To prevent the inadvertent operation, the motor controller fail-safe operation as already introduced for the main rotor propulsion system is implemented as well as an option for passivating a faulty motor output by implementing a power disconnect switch. 
The integration of a thermal management system and the information management system has been excluded so far, but will be included in further research.  

Based on the sizing of all components the total propulsion system weight for the presented multirotor excluding the thermal management systems is expected to reach 1144 kg as shown in Table \ref{tab:PropulsionSystemWeight}. An overview about further vehicle design parameters can be found within Table \ref{tab:ConceptParameters} of the appendix.

\begin{table}[]
\caption{Summary of the propulsion system weights} \label{tab:PropulsionSystemWeight}
\begin{tabular}{@{}p{1.8cm}ll@{}}
\toprule
                                & Weight per Unit [kg]   & Total Weight [kg]\\
\midrule
Main rotor propulsion system \tnote{1}    & 24.6 + 17 + 52     & 98.4 + 68 + 208\\ 
Push propulsion system \tnote{2}          & 12.3 + 8.5      & 24.6 + 17\\ 
Battery packs \tnote{2}                   & 120 / 248     & 480 + 248 \\
Total propulsion system                    &               & 1144  \\ 
\botrule
\end{tabular}
\footnotetext[1]{Weights indicated for motor weight,  motor controller weight and gearbox weight}
\footnotetext[2]{Main battery pack and push drive battery pack}
\footnotetext{Power distribution system and cooling system currently excluded from weight analysis.}
\end{table}

\FloatBarrier
\section{Conclusion and final evaluation}\label{sec:Conclusion}
Within this paper, initially a method was presented for the conceptual design of an \acrshort{evtol} propulsion system. This method was then applied to a multirotor vehicle for a specific intracity use case, with a special focus on developing a safe propulsion architecture, sizing each component and validating the architecture by simulation. 
In the following the results are structured to answer the initial research questions:
\vspace{5mm}

\textbf{How should  the  conceptual  design  process  of the  propulsion  system  be  carried  out  for  an all-electric  multirotor  \acrshort{vtol}  vehicle  that  is transporting passengers so that the safety goals of \acrshort{easascvtol} can be met?}\\
The conceptual design method as presented within section \ref{sec:Method} is divided in five Steps. Within step one the concept of operation needs to be defined, which includes defining the flight mission and payload requirements. Based on these requirements the vehicle configuration has to be preselected and the powertrain technology to be used is defined. Within step two several further requirements are developed which are based on the required controllability, the handling quality and allowed noise emission. Within the third step, the propulsion system is defined which can be segregated into defining the flight control system, the power and drive system, the electrical system and the thermal management system considering the previously established requirements. 
This propulsion system concept is then refined within the safety analysis and sized as well as validated within the vehicle sizing and simulation step. The system architecture refinement process is usually an iterative process between the concept definition, the safety analysis and the succeeding sizing step and is being conducted until the safety requirements of \acrshort{easascvtol} can be met. 
\vspace{5mm}

\textbf{What   is   the   impact   of   the   \acrshort{easascvtol}  reliablity  requirements  on  the  conceptual design of a multirotor propulsionsystem?}\\
Based on the safety \& reliability analysis it became apparent that the loss of one rotor lift and the incorrect operation of one rotor providing lift are the most critical system design drivers for the propulsion system of the main rotor in terms of the reliability requirements. Additionally, it was identified that, if the failure rate requirements can be fulfilled for those events, the other functional hazards will be fulfilled as well. Therefore, the focus of any system designer should be put on meeting the safety \& reliability requirement for the total and partial loss of one main rotor and for the incorrect operation of one rotor. 
Within the analysed case study, the safe propulsion system for the multirotor requires numerous redundancies. This includes that each main rotor requires two separate drive trains, which can be masked by two means in case of any malfunction within each drive train. One option for passivating the drive train is using a disconnect clutch, the other option is to cut the power to the electric motor by using a power disconnect relay. Additionally, each of the two drive trains driving the rotor must be designed for being capable to provide $200$ \% of normal power in case of any system malfunction. In case of a signal loss of the triple modular \acrshort{fcc} system, each motor controller must be designed as a dual active/passive module, which is additionally equipped with a backup-mode. This backup mode shall be able to set a constant rotor speed slightly below hover power for normal flight. Four battery packs are advisable to be used for the four main rotor propulsion system. Each motor controller should be connected to at least two batteries. Additionally, two push propellers are required which counteract the torque moment in case of one main rotor loss. 
The push-propeller architecture, is mainly driven by preventing the inadvertent operation of each propulsor. Consequently, the corresponding motor controllers should be connected to the fifth stand-alone battery pack and also incorporate a fail-safe backup mode. This time one disconnect relay is sufficient for passivating faulty drive outputs. 
In terms of the cooling system for each main rotor drive system it must be ensured, that not more than one cooling system fails simultaneously, as the failure of the cooling system results in the loss of the corresponding main rotor. 
For the cooling system design of the battery packs it must be ensured that not more than one battery pack is influenced in case of a cooling system failure. A complete loss of the cooling function may become a catastrophic event. 
\vspace{5mm}

\textbf{Which implications does an all-electric battery-powered \acrshort{evtol} have on the propulsion system architecture besides the safety requirements?} \\
Besides the safety analysis, the sizing and simulation process revealed the following: 
The sizing of the power and drive system identified that a gearbox is required for driving each main rotor of the quadcopter. Only by increasing the number of rotors, the gearbox could become obsolete.

During the sizing of the battery it became apparent that it is essential not only to check the energy and power requirement during normal operation, but also during emergency operation. The case study of this paper has shown, that the sizing of the main rotor battery packs for the presented architecture, is not driven by the energy amount or power requirement during normal operation but rather by the power requirement during emergency operation, in which two of four packs need to deliver 1.5 times the power compared to normal operation. The sizing of the push drive system battery pack is also driven by the emergency operation energy requirement which is 1.27 times the energy requirement for normal operation. 
\vspace{5mm}

\textbf{Which requirements must be met by a thermal management system of the developed all-electric multirotor propulsion system?} \\
A battery-electric multirotor propulsion system requires at least two cooling circuits.
The sizing of the thermal management system for the power and drive components revealed that the electric motors and the motor controllers of each drive unit can be cooled within the same liquid cooling system, whereas the battery requires a separate cooling system due to the different operating temperatures. For cooling the electric motor and motor controller it is not sufficient to rely on the airstream. However, a combined cooling of airflow cooling and liquid cooling should be preferred. The liquid cooling circuit can cool the electric motor as well as the corresponding motor controllers of one drive unit simultaneously, by connecting them, for example, in series. This exemplary liquid cooling circuit requires at least 0.14 kg/s flow rate, using Glysantin G40. This enables to keep the electric motor below 90 °C during normal operation and just below 120 °C during emergency operations, even at ambient temperatures up to 42.7 °C. Additionally, the heat exchanger and the electric motors must be placed within the airstream to allow for additional air cooling. The heat exchanger can be expected to weight around 4 kg with a size of $0.3\cdot0.3\cdot0.3$ m. In order to cool down the heated components after each flight it is necessary to keep the cooling system operative on ground as well. This cool down can take up to 30 minutes depending on the outside temperature. 
A secondary liquid cooling circuit is required in order to keep the battery packs below 40 °C operating temperature, as without any cooling circuit the batteries would heat up above 40 °C even at ambient temperatures of 20 °C. A cooling circuit using a glycol water mixture of 20:80 with a mass flow of 0.0695 kg/s is expected to keep the operating temperature of the battery pack within $T_{amb} + 5\;^{\circ}C$ during normal operation and in case of an emergency procedure within $T_{amb} + 7\;^{\circ} C$. The liquid cooling system however, can only be used up to ambient temperatures of 36.2 °C. Higher ambient temperatures require a refrigeration circuit. In order to ensure a minimum battery operating temperature of 20 °C a heating circuit is advisable as soon as the ambient temperature falls well below 20 °C.

Comparing the results to the literature identified within section \ref{sec:StateOfArt}, specifically the research of \citet{Darmstadt.2021}, this work presents an alternative solution for a safe propulsion system design for a quadcopter. In addition, the implications of such a propulsion system on the total propulsion system mass and a validation of the system architecture based on current technology is provided through simulation models. 

\section{Future perspective}\label{sec:FuturePerspective}
Within this research a conceptual design process for achieving a safe propulsion system for \acrshort{evtol} multirotors was presented.
As the focus was set primarily on designing a reliable drive system, other system groups that are linked to the propulsion system, like the information management system, the electrical system, the thermal management system as well as safety systems need further investigation. 
Firstly, the thermal management system as well as the information system group need to be included in future safety analysis. 
Secondly, the electrical system architecture, therein especially the power distribution, should be analysed in further detail. As battery degradation has so far not been considered, its effect on the sizing of the battery packs should be analysed as well. 
Thirdly, the safety system requirements defined within the \acrshort{easasce19} \cite{EASA.SCE19} have to be included within the propulsion architecture design which includes e.g. means to prevent and cope with uncontrolled fire within the battery system. 
Thirdly, the rotor, rotor shaft connection as well as the junction between the two gearboxes and the rotor shaft need to be designed from a mechanical perspective and investigated to prevent single point of failures. 
Forth, the thermal management system of the battery needs to be extended as the currently assumed liquid cooling circuit is only able to provide sufficient cooling below ambient temperatures of 36.4 °C. Therefore, a lightweight and safe refrigeration circuit should be assessed as an alternative. In order to ensure the correct battery operating temperature even at low ambient temperatures, adding a heating possibility to the battery pack thermal management system probably using the heat of the motor and motor controllers should be considered. Additionally, a comparative study for the liquid cooling circuit of the electric motors and motor controllers should be conducted to assess the implications of a two-step cooling instead of the currently evaluated one-step cooling system.
Fifth, the heat development within the power and drive system of the rear propulsion needs to be investigated and the thermal management system adapted accordingly.
As the presented propulsion architecture is only valid, if the vehicle can continue safe flight and landing even during a single rotor loss, further investigation is required to establish corresponding means of controlling such a flight state.

\subsection{Contribution of this work towards minimizing costs and maximizing benefits of a UAM system}
\begin{figure}[htpb]
    \centering
    \includegraphics{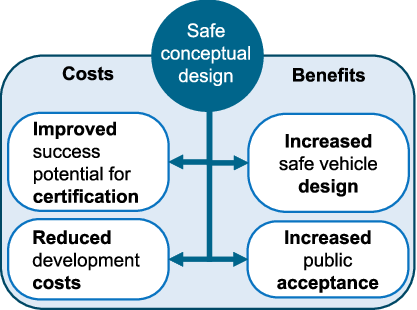}
    \caption{Mapping the areas of the UAM overall system that are positively impacted by the presented work.}
    \label{fig:WorkContributionUAMSystem}
\end{figure}
In accordance with the leitmotif "Opportunities and Challenges of Urban Air Mobility" this section evaluates how this work contributes to advancements within the \acrfull{uam} system (see Fig. \ref{fig:WorkContributionUAMSystem}). The contribution in making \acrshort{uam} become reality can be grouped in minimizing \acrshort{uam} costs or maximizing \acrshort{uam} benefits. As this work presented a method for the conceptual safe design of the propulsion system as well as its implications on the propulsion system architecture for an exemplary \acrshort{uam} concept of operation, this paper primarily adds value towards increasing the reliability of a vehicle design. By providing means for a model-based systems engineering approach for the safe vehicle design, the chance of a fast and successful certification process may also be increased. Additionally, by taking safety aspects into account already during the conceptual design phase, subsequent high vehicle development costs due to late design adjustments can be prevented. On the other hand, the safe propulsion design as presented for the multicopter may positively influence the passenger's acceptance towards these vehicles. 
\backmatter

\section*{Statements and Declarations}
\bmhead{Author Contributions}\mbox{}\\
Conceptualization: F.J.; Methodology: F.J., O.B.; Writing - original draft preparation: F.J., O.B., S.M.L., A.H.B., J.R., L.B.; Writing - review and editing: F.J.; Supervision: O.B.; project administration, O.B.;
All authors have read and agreed to the published version of the manuscript. 

\bmhead{Acknowledgments}\mbox{}\\
The paper presents results from the internal DLR project HorizonUAM.

\bmhead{Competing Interests}\mbox{}\\
The authors have no competing interests to declare that are relevant to the content of this article.
\noindent

\bigskip


\newpage
\section*{Appendix A}\label{secA1}

\begin{table}[htpb]
\caption{Summary of the HorizonUAM multirotor vehicle concept parameters} \label{tab:ConceptParameters}
\begin{tabular}{@{}ll@{}}
\toprule
                        & Multirotor concept vehicle  \\
\midrule
Propulsion Type                     & Battery-Electric \\
Control Scheme                      & Rotor Speed Control \\
Design Gross Weight                 & 1954 kg\\
Payload Capacity                    & 360 kg\\
Number of Rotors                    & 4 \\
Rotor Radius                        & 2.64 m            \\
Max Rotor Tip Speed                 & M.45                     \\
\hspace{1mm}(Normal Ops)            &  \\
Max Rotor Tip Speed                 & M.65                      \\
\hspace{1mm}(Emergency Ops)         &  \\
Number of forward                    & 2\\
\hspace{1mm} facing propellers     &   \\
Propeller Radius                    & 0.54 m\\
Max Propeller Tip Speed             & M.54 \\
Battery Capacity                    & 263 Ah / 159 Wh\\
Design Range                        & 3x16.7 km + 20 min Loiter\\
\botrule
\end{tabular}
\end{table}

\FloatBarrier




\end{document}